\documentclass[%
 reprint,
nofootinbib,
 amsmath,amssymb,
 aps,
]{revtex4-2}
\usepackage{makecell}
\usepackage{amssymb}
\usepackage{xcolor}
\usepackage{graphicx}
\usepackage[pdfborder={0 0 0},colorlinks=true,linkcolor=blue,urlcolor=blue,citecolor=blue]{hyperref}
\hypersetup{
           breaklinks=true,   
           colorlinks=true,   
           pdfusetitle=true,  
        }
\usepackage{dcolumn}
\usepackage{bm}

\newcommand{\orangetxt}[1]{{\color{blue} #1}}
\begin{document}

\preprint{APS/123-QED}

\title{Virialized Profiles and Oscillations of Self-interacting Fuzzy Dark Matter Solitons}

\author{Milos Indjin}
 \homepage{m.indjin@newcastle.ac.uk}

\author{I-Kang Liu}
\homepage{i-kang.liu1@newcastle.ac.uk}
\author{Nick P. Proukakis}
\homepage{nikolaos.proukakis@ncl.ac.uk}
\author{Gerasimos Rigopoulos}
\homepage{gerasimos.rigopoulos@ncl.ac.uk}

\affiliation{
School of Mathematics, Statistics \& Physics, 
Newcastle University, 
Newcastle upon Tyne, 
NE1 7RU 
UK
}
\date{\today}

\begin{abstract}

We investigate the effect of self-interactions on the shape and oscillations of the solitonic core profile of condensed fuzzy dark matter systems without the backdrop of a halo, revealing universal features in terms of an appropriately scaled interaction strength characterizing the crossover between the weakly- and strongly-interacting regimes. Our semi-analytical results are further confirmed by spherically symmetric simulations of the Gross-Pitaevskii-Poisson equations. Inverting our obtained relations, we highlight a degeneracy that could significantly affect constraints on the boson mass in the presence of repulsive boson self-interactions and propose the simultaneous extraction of static and dynamical solitonic features as a way to uniquely constrain both the boson mass and self-interactions.

\end{abstract}

\maketitle

\section{Introduction}

Fuzzy dark matter (FDM) has lately emerged as a popular alternative to the conventional cold dark matter (CDM) model. Following the first proposition of an ultralight bosonic dark matter constituent \cite{Hu2000}, an ultralight particle within the mass range $10^{-22} - 10^{-20}$ eV, leads to a description based not on incoherent particles, but on a single coherent wavefunction, with a de Broglie wavelength of $\sim$ $\mathcal{O}(1)\,\, \mathrm{kpc}$. Numerical simulations \cite{Schive2014, Mocz2017, May2021StructureDynamics} have shown that on large scales, FDM replicates the cosmic web structure of cold dark matter, whilst addressing some alleged challenges that CDM faces on galactic scales. Formation and growth of structures in such a dark matter background has been thoroughly investigated recently~\cite{Mocz2019, Mocz2020, Banerjee2020, Mina2022}. The favourable look upon this type of dark matter in recent years has been motivated by the inherent properties which remedy ailments from which the CDM model allegedly suffers \cite{2015PNAS..11212249W, 2017ARA&A..55..343B, 2017Galax...5...17D}. Among these, most notably, is the cusp-core problem \cite{Maccio1} which is resolved in FDM since the balance between the quantum pressure and gravitation in FDM naturally forms a core, embedded in a NFW-like halo~\cite{Rindler-Daller2014}. Furthermore, some tentative, more direct favourable evidence for FDM may have begun to emerge, see \textit{e.g.}~\cite{Amruth2023}. Scrutiny of this model has become an active area of research - see \textit{e.g.}~\cite{2016PhR...643....1M, 2021ARA&A..59..247H, 2021A&ARv..29....7F} for recent reviews and references to the literature.

The cores in FDM halos, often termed FDM solitons, exhibit interesting dynamical behaviours. Given non-symmetric initial conditions, the cores formed from gravitational collapse exhibit a random walk within the base of the gravitational potential \cite{Li2021, Hui2021-2}. Furthermore, scalar field oscillations are observed in numerical simulations \cite{Woo1, Schive2014, Mocz2017, Li2021, Marsh2019, Veltmaat2018, Hui2021-2, Glennon2020-2}, and significant advances have been made in understanding the parameter dependence of their frequency \cite{Chavanis2011, Chavanis2011_2}. These oscillations manifest in the form of homogeneous radial expansion and contraction of the soliton core, and have been found to send out density waves into the surrounding halo \cite{Woo1}. Higher excited asymmetric states are also possible but they will not be discussed in this paper. In such non-symmetrically initialised simulations, the constructive interference of density waves and further stochastic density fluctuations \cite{Church2019, Chiang2021-2} also generate granules with a typical scale of the order of the central soliton core \cite{Schive2014,Schive2014a,Mocz2017}. These granules have been stipulated to be the driving force behind the dynamical heating of stellar streams in the Milky Way \cite{Amorisco2018}, whilst density waves have been proposed as potential sources of disk heating \cite{Church2019}. Although the relation and interactions of the solitons to their host halos are very interesting and still to be fully explored, in this paper we will examine the FDM soliton in isolation, without the backdrop of a complete halo. 

Most FDM investigations have considered non-interacting bosons, for which the physical state of the system is described by a Schr\"{o}dinger equation, coupled to the Poisson equation, in what is known as the Schr\"{o}dinger-Poisson system of coupled equations (SPE)\footnote{In this paper we refer to both the non-interacting, standard, fuzzy dark matter and its generalized form including self-interactions as FDM, with the presence or absence of interactions being clear from the context.}.
However, lacking any evidence to the contrary, the addition of particle-particle interactions renders the situation more interesting and dynamically complex -- such a set-up, based on contact interactions adding a further nonlinear contribution to the Schr\"{o}dinger equation, has been the subject of many recent works\footnote{For earlier related work in a general relativistic context, see \textit{e.g.}~\cite{Khlopov1985, Bianchi1990} and citing references.}~\cite{Chavanis2011,Chavanis2011_2, Rindler-Daller:2011afd, Li2014, Li2017, Desjacques:2017fmf, Chavanis:2020rdo, Hartman2022,Hartman2022-2,Mocz_2023, Chakrabarti2022, Dave2023}. Although some works focus on the limiting case when interactions dominate (and the kinetic energy generating the quantum pressure becomes negligible), \textit{e.g.}~\cite{Dawoodbhoy2021, Shapiro2021, Rindler-Daller2022}, effort has also been made to obtain general behaviours of self interacting FDM systems~\cite{Chavanis2011, Chavanis2020}. Furthermore, constraints upon the possible strength of self-interactions along with the boson mass have very recently been examined ~\cite{Li2014, Li2017, Delgado2022, Hartman2022, Chakrabarti2022} and, as a result, typical values for a repulsive self-coupling of $g \sim (10^{-32} - 10^{-26})\, \mathrm{Jm^{3}kg^{-1}}$ are usually quoted within the context of interacting FDM literature, with the exact value depending on the chosen boson mass. Even in the lower ranges, such a small self-interaction strength can still lead to clear and notable effects, which could be observationally relevant, with ~\cite{Delgado2022} presenting favourable observational evidence for the existence of a non-zero self-coupling. Moreover, although attractive FDM solutions are typically unstable beyond a small range of allowed values, potentially leading to collapse and/or the formation of a black hole~\cite{Chavanis2011, Chavanis2016}, recent work~\cite{Mocz_2023} highlights the potentially favourable and desired role of attractive interactions in enhancing small-scale structure formation in large simulations of the cosmic web. Such examples of self interacting FDM would leave observational signatures like an increased number of solitons that form in areas of constructive interference in cosmic filaments, where instabilities are present under the attractive self interaction.

The aim of the present work is to characterize in detail -- both analytically and numerically -- the effect of repulsive interactions on the size and shape of the underlying virialized solitonic core, and to use such information to accurately predict the oscillations of the core -- building on important earlier work on such observables~\cite{Veltmaat2018, Guzman2019, Schive2020, Veltmaat2020-Baryons, Chiang2021-2, Chowdhury2021}.
Based on this we propose a scheme for uniquely identifying both the self-interaction strength and the boson mass by combining information of soliton shape details and oscillation frequency, which could potentially become available in future observational data analysis. Indeed, oscillations perturb the rotational velocity curves significantly within the soliton and in the region immediately adjacent to it~\cite{Chowdhury2023}. A statistical evaluation of the gravitational heating of galaxies could therefore give a preliminary indication of a relationship between frequency and other soliton parameters.

Firstly, building upon earlier work based on a Gaussian solitonic core of some interaction-dependent effective width~\cite{Chavanis2011}, and making use of the virialization condition, we derive analytical expressions characterizing the change in shape parameters as a function of interaction strength in the context of the empirical model of solitonic cores~\cite{Schive2014,Mocz2017,Chan2018,Chiang2021-2}. 
Further introducing a characteristic interaction strength marking the crossover between weakly-interacting and strongly-interacting regimes -- defined by equating the interaction energy to the quantum kinetic energy (see also \cite{Chavanis2011, Chavanis:2020rdo, Rindler-Daller2022}) -- we are able to obtain analytical results for appropriately scaled soliton parameters (radius, peak density), cast only in terms of the ratio of the interaction strength to such characteristic value; such results smoothly interpolate between the previously studied non-interacting and strongly-interacting (Thomas-Fermi) limits. 

To test the accuracy of our extended analytical predictions, we also perform detailed spherically symmetric numerical simulations of isolated soliton cores generated through the very efficient imaginary time propagation method widely used in cold atom studies~\cite{Minguzzi-Imaginary-Time} (but scarcely seen in cosmological works - see however \cite{Chavanis2020-2, Dmitriev2021}).
We compare and contrast our virialized solitonic shapes both in the absence and presence of repulsive self-interactions to the Gaussian~\cite{Chavanis2011}, Thomas-Fermi~\cite{Chavanis2011} and empirical~\cite{Schive2014} profiles, with our findings providing a direct bridge between all such limiting cases.

Moreover, our scaled analytical formulae are numerically confirmed to be independent of the values of the boson mass ($m$) and total soliton mass ($M$), at least within the probed ranges $10^{-22} \mathrm{eV/c^2} \leq m \leq 10^{-20} \mathrm{eV/c^2}$ and $10^6 M_\odot \leq M \leq 10^8 M_\odot$.
Such an analysis facilitates a critical test of the typically-used empirical profile, and provides a direct relation of the scaling of central core density and soliton width as a function of scaled interaction strength, smoothly interpolating between relevant previous findings.

Next we evaluate the oscillation frequency of the core, building upon the early analysis of Chavanis \cite{Chavanis2011}, but based here not on a Gaussian approximation to the soliton, but on the more accurate empirical profiles for which we
explicitly derive a relationship between the peak density, $\rho_0$, and the soliton oscillation frequency, $f$ -- thus supplementing previous related works based solely on numerical fitting~\cite{Veltmaat2018, Guzman2019, Schive2020, Veltmaat2020-Baryons, Chiang2021-2, Chowdhury2021}. 
Implementing a perturbative approach on our accurately virialized numerically generated solitons, we obtain accurate numerical predictions for these oscillations which we contrast to our analytical predictions, and simpler previous ones.

Our numerical analysis of soliton shape (peak density, width) and oscillation frequency point to the existence of a degeneracy between boson mass and self interaction strength in determining the resulting soliton parameters. We discuss the implications of such findings for uniquely determining both interaction strength and boson mass from potential observational signatures. Through this analysis, we highlight that care is necessary when imposing constraints on the boson mass in FDM, as the addition of a free parameter in the self-interaction strength could significantly widen the allowed parameter space for the boson mass, as previously suggested in~\cite{Chavanis:2020rdo}. We thus indicate that strong constraints such as those discussed in~\cite{Marsh2019, Chiang2021-2} should likely be revisited with this in mind. 

The paper is structured as follows:
After reviewing the governing equations of FDM and common soliton profile shapes discussed in the literature (Sec.~\ref{sec:background}), we critically revisit the shapes of virialized soliton profiles (Sec.~\ref{Sec:virialized_profiles}). Here we review the non-interacting results (Sec.~\ref{sec:vir_cons}), and focus on the modifications imposed by repulsive self-interactions, which provide new extended relations in terms of a characteristic interaction strength approximately separating the weakly- and strongly-interacting limits: this is obtained by matching quantum kinetic and interaction energies (Sec.~\ref{Sec:SI}). Our numerical procedure and results are then shown (Sec.~\ref{Sec:Numerical Approach}), with
soliton oscillations discussed both analytically and numerically (Sec.~\ref{Sec:soliton_oscillations}). 
By inverting relations for soliton core density and oscillation frequency parameters previously obtained in terms of a fixed boson mass $m$, we obtain implicit relations for $m(g)$ as a function of variable interactions for any radius/oscillation frequencies (Sec~\ref{Sec:degeneracy}), pointing to emerging degeneracies, whose inferred observational constrain implications are discussed (Sec.~\ref{sec:observations}). Brief conclusions are then given in Sec.~\ref{Sec:conclusions}.

\section{\label{sec:background} The Gross-Pitaevskii-Poisson system }

In the framework of FDM, dark matter is composed of ultra-light bosonic particles and can be described by a classical complex field $\psi(\mathbf{r},t)$. We will take its \emph{mass density} to be represented by $\rho = |\psi|^2$ and the energy functional for the field can be written as
\begin{eqnarray}
E &=& \int d^3 r \left[ \frac{\hbar^2}{2m^2}|\nabla \psi|^2 + \frac{1}{2}V[\psi]\,|\psi|^2 + \frac12 \frac{g}{m} |\psi|^4 \right] \nonumber \label{eq:energy_functional}\\
 &\equiv& \Theta + W + U
\end{eqnarray}
where $\Theta$, $W$ and $U$ are the kinetic energy, the gravitational energy and the interaction energy respectively. The Newtonian potential V(\textbf{r},t) is a functional of $\psi$ and obeys the Poisson equation,
\begin{equation}
    \nabla^2 V(\mathbf{r},t)  = 4\pi G \left( \rho - \bar{\rho}\right)\label{eq:Poisson}
\end{equation}
where the average mass density, $\bar{\rho}$, has been subtracted as is necessary for the consistent description of continuous systems under Newtonian gravity \cite{Chavanis2011, Liu2023}. The self-interaction strength, $g$, is considered as a two-body contact (short-range) interaction which can be effectively associated with the $s$-wave scattering length $a_s$ through the relationship  $g=4\pi\hbar^2a_s/m^2$.

By varying Eq.~(\ref{eq:energy_functional}), one arrives at a non-linear Schr\"odinger equation, a Gross-Pitaevskii type equation of the form
\begin{equation}
    i\hbar\frac{\partial \psi(\mathbf{r},t)}{\partial t} =\left[ -\frac{\hbar^2\nabla^2}{2m} + 
g\rho(\mathbf{r},t)+mV(\mathbf{r},t) \right]\psi(\mathbf{r},t)  \label{eq:GPPE} \;.
\end{equation}
which, along with equation (\ref{eq:Poisson}) constitute the Gross-Pitaevskii-Poisson Equation system (GPPE) for the description of FDM. By applying the Madelung transformation, $\psi=\sqrt{\rho}\exp(iS)$, and separating real and imaginary parts, we obtain the continuity equation,
\begin{equation}
    \frac{\partial \rho}{\partial t} + \nabla \cdot \left(\rho \bf{u}\right) = 0, \label{eq:continuity}
\end{equation}
in which $\mathbf{u}=\hbar\nabla S/m$ defines the velocity and the equation of motion takes the form of the generalised Bernoulli equation
\begin{equation}
    \frac{\partial S}{\partial t} + \frac{1}{2m}\left(\nabla S\right)^2 + mV + g\rho - \frac{\hbar^2}{2m}\frac{\nabla^2 \sqrt\rho}{\sqrt\rho} = 0. \label{eq:bernoulli}
\end{equation}
The last term of Eq.~(\ref{eq:bernoulli}) is the quantum pressure, which originates from the uncertainty principle. In the non-interacting case ($g = 0$), equilibrium is achieved when the repulsion due to the quantum pressure balances the attraction due to gravity \cite{Membrado1989}.

The virial theorem in this case states that~\cite{Guzman:2006yc,Chavanis2011}
\begin{equation}\label{eq:virial full}
    2 \Theta + W + 3U = 0.
\end{equation}
It is customary to decompose the kinetic energy into classical and quantum parts, respectively defined by
\begin{equation}
    \Theta_C = \int d^3\mathbf{r} \, \frac{1}{2}\rho|\mathbf{u}|^2, \quad  \quad \Theta_Q = \int d^3\mathbf{r} \, \frac{\hbar^2
    }{2m^2}|\nabla \sqrt{\rho}|^2\,.
\end{equation}
In equilibrium $\mathbf{u}=0$, and therefore the classical kinetic energy is zero, but the quantum kinetic energy, $\Theta_Q$, still contributes to the energy budget. The equilibrium system is static and obeys the virial theorem where only the quantum kinetic energy plays a part, thus
\begin{equation}\label{eq:virial}
    2 \Theta_Q + W + 3U = 0.
\end{equation}

\subsection{Common Soliton Profiles}
A characteristic feature of the Gross-Pitaevskii-Poisson system is the coherent core-like soliton structure at the centre of the gravitational well of the halo~\cite{Rindler-Daller2014,Marsh2019, Zagorac2022, Li2021}.
We discuss below characteristic expressions that are frequently used in the literature to describe the spatial profile of the soliton in various limits.

Firstly, in the absence of self-coupling ($g=0$), the gravitational attraction is solely balanced by the quantum pressure term. In this limit, the soliton density is usually approximated by the empirical profile~\cite{Schive2014}
\begin{equation}
    \rho = \rho_0\left( 1+\lambda\left( \frac{r}{r_c} \right)^2 \right)^{-8}
    \label{eq:emp_prof}
\end{equation}
where $\rho_0$ is the central (peak) core density and  $r_c$ represents the core radius, defined as the radius at which $\rho(r_c)=\rho_0/2$: this yields the exact value $\lambda=2^{1/8}-1\approx0.091$. We note here that while the above profile automatically 
satisfies the virial theorem $2\Theta_Q + W = 0$, in order to ensure this it is in fact critical to use the exact (as opposed to the approximate) value of $\lambda$ quoted above (see also subsequent related comments).
 
  We also stress here that $\rho_0$ and $r_c$ are not independent parameters, and a central density-radius relation has been reported~\cite{Schive2014,Schive2014a, Veltmaat2020-Baryons, Chiang2021-2, Chowdhury2021} in the form 
\begin{equation}
    r_{c}=\left(\frac{\rho_{0}}{1.9 \mathrm{M}_{\odot} \mathrm{pc}^{-3}}\right)^{-1 / 4}\left(\frac{m}{10^{-23} \mathrm{eV}}\right)^{-1 / 2} \mathrm{kpc} \;.
    \label{eq:emp_radius}
\end{equation}
Such an expression can also be obtained from variational energetic considerations~\cite{Liu2023}.

Secondly, the opposite, strongly-interacting limit, corresponds to a hydrostatic equilibrium in which the gravitational attraction is balanced by repulsive interactions~\cite{Chavanis2011}. In this limit, the solitonic core is instead approximated by a Thomas-Fermi profile~\cite{Lee1996, Goodman2000, Arbey2003, Bohmer2007}
\begin{equation}
     \rho_{TF}(r) =\rho^{TF}_0 \left( \frac{ R_{TF}}{\pi r} \right) \sin\left(\frac{\pi r}{R_{TF}}\right)\,,
\end{equation}
of characteristic spatial extent $R_{TF}$.

For consistency with the notation used for the empirical profile, we introduce here the spatial extent, $r_c^{TF}$, corresponding to the point where the density drops to half its central value, i.e.~$\rho_{TF}(r_c^{TF})=(1/2) \rho_{TF}(r=0)$, thus yielding
\begin{equation}
     \rho_{TF}(r)=  \frac{\rho^{TF}_0 }{1.895 }\frac{r_c^{TF}}{r}\sin\left(1.895\frac{r}{r_c^{\small TF}}\right) \;. \label{eq:TF_prof}
\end{equation}

More generally, but only accounting for global shape features without attention to detailed radial dependence, the soliton can also be approximated by a Gaussian~\cite{Chavanis2011}: while only approximate, and a poor description for both the $g=0$ and $g \rightarrow \infty$ limits, such description has two obvious benefits: 
(i) it facilitates an easy qualitative interpolation across these two limiting cases in the context of an effective interaction-dependent width, and
(ii) it is easily amenable to analytical calculations.
The form of such a Gaussian is
\begin{equation}
    \rho(r) = M\left(\frac{1}{\pi r_c^2}\right)^{3/2} \exp\left( -\frac{r^2}{r_c^2}\right). \label{eq:gaussian}
\end{equation}

\begin{figure}
    \centering
    \includegraphics[width=1\linewidth,keepaspectratio]{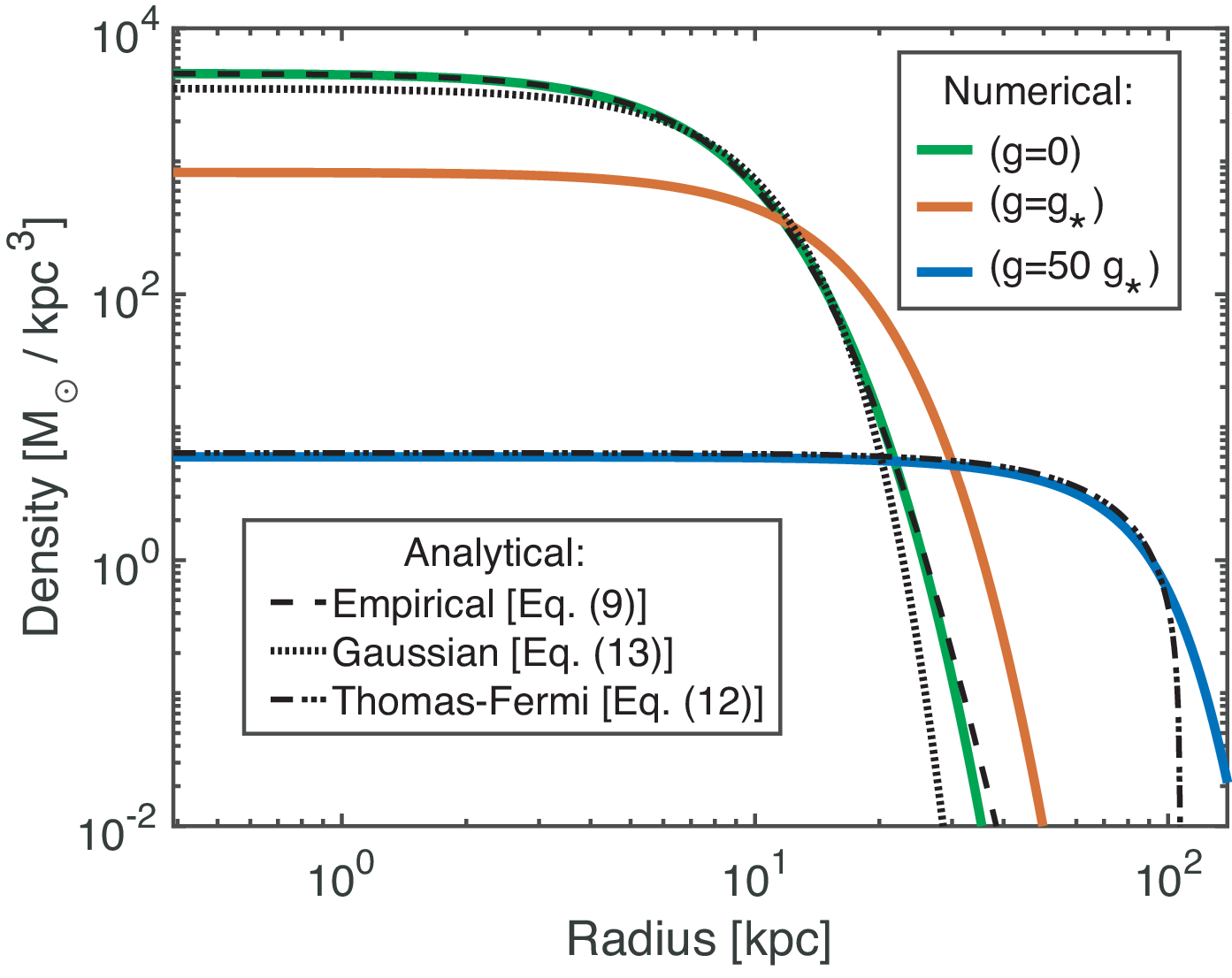}
    \caption{ 
    Dependence of the shape of density profiles on interaction strength for characteristic examples, and comparison between numerical (generated using imaginary time propagation to find the ground state solution) and analytical predictions in such regimes, with interaction strength increasing from top to bottom curves.
Compared to the non-interacting limit (top green line), we also show a profile with self-interactions of an adequate strength ($g_*$) to impact the properties of the soliton (orange line), and the corresponding profile in the Thomas-Fermi regime of strong self-interactions.
The empirical profile is given in Eq.~(\ref{eq:emp_prof}), the  Thomas-Fermi and Gaussian and profiles are respectively defined by  Eq.~(\ref{eq:TF_prof}) and Eq.~(\ref{eq:gaussian}).
In all cases, the profiles are normalised to $M = 1 \times 10^7 ~M_\odot$, and with $m = 2 \times 10^{-22} ~\rm{eV}/\rm{c}^2$. 
The characteristic interaction strength $g_*$ corresponds to the value marking an approximate crossover from weakly-interacting to strongly-interacting, and is defined in Eq.~(\ref{eq:g_*}) by balancing interaction and quantum kinetic energies -- see subsequent discussion in the text.
    }
    \label{fig:profComp}
\end{figure}

The different arising density profiles and a comparison with our more accurate numerical results (discussed in Sec.~\ref{Sec:Numerical Approach}) are shown in Fig.~\ref{fig:profComp} across the entire range of interactions, with $g_*$ denoting a characteristic  interaction strength (defined later) marking an approximate crossover from the non-interacting to the strongly-interacting limits. 

\subsection{Generalised Soliton Profile Ansatz \label{sec:soliton_ansatz}}

In order to more accurately characterize the role of interactions on the soliton shape, we proceed by constructing a generalized ansatz.
On physical grounds, 
we anticipate for such a general density profile to depend on the radial coordinate $r$, and the interaction strength $g$, only through the dimensionless ratios $r/r_c$ and $g/g_*$, where $g_*$ is a characteristic interaction strength to be defined later.
We thus introduce a generalised ansatz of the form \footnote{See \cite{Chavanis:2020rdo} for a general density profile ansatz and \cite{Rindler-Daller2022, Chavanis2011} for a similar scaled self-interaction strength.}
\begin{equation}
    \rho = \rho_0 \, \varphi\left(\frac{r}{r_c}, \Gamma\right), \label{eq:general_ansatz}
\end{equation}
where 
\begin{equation}
\Gamma = \frac{g}{g_*}. \label{eq:gamma}
\end{equation}

Following Chavanis~\cite{Chavanis2011} we start our analytical studies by reducing the expression for the mass and various system energies in terms of relevant physical variables and appropriate dimensionless integrals, whose values can be calculated based on our generalised ansatz. Specifically we obtain the following expressions:\\
\\
Total Mass:
\begin{align}
    M &= 4\pi \int_0^\infty \rho_0 \, \varphi\left(\frac{r}{r_c}, \Gamma\right) r^2 dr, \\  \label{eq:Mass}
      &=\eta(\Gamma) \, \left(4\pi \rho_0 r_c^3\right).
\end{align}
Moment of Inertia:
\begin{align}
    I &= 4\pi \int \rho_0 \, \varphi\left(\frac{r}{r_c}, \Gamma\right) r^4 dr, \\
      &= \alpha(\Gamma)\, \left(M r_c^2\right).
\end{align}
Quantum Kinetic Energy:
\begin{align}
    \Theta_Q &= \frac{2\pi \hbar^2
    }{m^2} \int \,  \left| \frac{\partial}{\partial r} \left[\rho_0 \, \varphi\left(\frac{r}{r_c}, \Gamma\right)\right]^{1/2}\right| ^2 \, r^2\, dr\\ \label{eq:Q_kin}
    &=  \sigma(\Gamma)\, \left(\frac{\hbar^2 M}{m^2 r_c^2}\right).
\end{align}
Gravitational Energy:
\begin{align}
W &= \frac{8\pi G}{2} \int_0^\infty r M(r) \rho_0 \, \varphi\left(\frac{r}{r_c}, \Gamma\right) dr, \\ \label{eq:W_grav}
  &= -\nu(\Gamma)\, \left(\frac{G M^2}{r_c} \right). 
\end{align}
Interaction Energy:
\begin{align}
    U &= \frac{2\pi g}{m} \int \rho_0^2\,  \varphi^2\left(\frac{r}{r_c}, \Gamma\right) r^2 dr, \\
    &= \zeta(\Gamma)\, \left(\frac{M^2 g}{2mr_c^3}\right). \label{eq:interaction_energy}
\end{align}

In the above expressions, $\eta(\Gamma)$, $\alpha(\Gamma)$, $\sigma(\Gamma)$, $\nu(\Gamma)$ and $\zeta(\Gamma)$ are parameters depending on the dimensionless interaction strength, obtained from a dimensionless integral directly related to the shape of the density profile of the soliton in the presence of self-interactions. We shall henceforth refer to such parameters as shape parameters. Such parameters were first introduced in the pioneering work of Chavanis~\cite{Chavanis2011} who used a Gaussian approximation to the soliton density profile, given by Eq.~(\ref{eq:gaussian}). We calculate the specific shape parameters for the three aforementioned profiles in Table~\ref{table:constants}. 

As evident, the values of these shape parameters are rather sensitive to the exact profile shape.
Moreover, we  emphasize that calculations of soliton static and dynamical parameters (e.g.~radius, peak density, oscillation frequency) will be found to exhibit a high degree of sensitivity in changes to the shape of the profile -- and thus the shape parameters -- already in the non-interacting limit, an effect amplified when interactions are taken into consideration.

\begin{table*}[]
\begin{tabular}{|c|c||c|c|c|c|c|}
\hline
\makecell{Physical \\ Origin} & \makecell{Shape \\ Parameter} & \makecell{Gaussian \\ ~[Eq.~(\ref{eq:gaussian})]} & \makecell{Empirical \\~[Eq.~(\ref{eq:vir_profile})]} & \makecell{Numerical \\ $g=0$} & \makecell{Numerical \\ $\Gamma \in $ $\left[ 10^{-5},~10^4\right]$} & \makecell{Thomas-Fermi \\~[Eq.~(\ref{eq:TF_prof})]} \\ \hline \hline

Mass & $\eta$   & 0.4385 &  0.9347 & 0.9693 $\pm 0.0231$ & \orangetxt{0.4731} - 1.0270 & \orangetxt{0.4385}\\ \hline
Mom.~of Inertia & $\alpha$ & 1.5000    & 3.0009 & 3.0531 $\pm 0.0578$ & \orangetxt{1.1163} - 3.1435& \orangetxt{1.0497}\\ \hline
Quant.~Kin.~Energy & $\sigma$ & 0.7500 & 0.3919 & 0.3865 $\pm  0.0038$ &  0.3799 - \orangetxt{1.3053}  & \orangetxt{1.2266}\\ \hline
Grav.~Energy & $\nu$    & 0.3989 & 0.2919 & \,\, 0.3033 $\pm 0.0158$ \,\, & 0.2936 - \orangetxt{0.4482} & \orangetxt{0.4584}\\ \hline
Int.~Energy & $\zeta$  & 0.0635 & 0.0262 & 0.0247 $\pm 0.0007$ & 0.0235 - \orangetxt{0.0823} & \orangetxt{0.0896}  \\ \hline
\end{tabular}
\caption{
Values for the various shape parameters arising  from the relevant energy integrals depending on the exact profile shapes: Such values are obtained analytically for the cases of a Gaussian profile (an analysis already performed in Ref.~\cite{Chavanis2011} from which the values in the 3$^{\rm rd}$ column are quoted), and evaluated in this work also analytically for the empirical profile of Eq.~(\ref{eq:vir_profile}) (4$^{\rm th}$ column). These are compared and contrasted to results from the numerical simulations in the non-interacting regime $g=0$ (5$^{\rm th}$ column), and over a broad range of relevant interaction strengths smoothly interpolating between the non-interacting and strongly-interacting limits, showing the change in the shape parameters as a function of interactions (penultimate column).
To highlight the observation that the values of some shape parameters decrease ($\eta$, $\alpha$), while others ($\sigma$, $\nu$, $\zeta$) increase with increasing interaction, we depict the limiting high interaction strength values by blue. For comparison, the final column shows the analytically predicted values in the strongly interacting Thomas-Fermi regime, based on the Thomas-Fermi profile of Eq.~(\ref{eq:TF_prof}). A graphical representation of the dependence of such shape parameters on scaled interaction strength $\Gamma$ is shown in Fig.~\ref{fig:allCoeffs}.
}
\label{table:constants}
\end{table*}

\section{Virialized Soliton Profiles } \label{Sec:virialized_profiles}

\subsection{Non-Interacting Virialized Profiles ($g=0$)\label{sec:vir_cons}}

In the absence of interactions ($g=0$),
the Virial theorem between the quantum kinetic and the gravitational energies 
\begin{equation}\label{eq:Virial2}
    2 \Theta_Q + W = 0
\end{equation}
can be used, jointly with the above energy expressions 
to establish a direct relationship between peak density and core radius \cite{Chavanis2011}. 

To contrast the non-interacting limit of this section to subsequent considerations featuring the additional inclusion of interactions, we
henceforth label all such $g=0$ (i.e.~$\Gamma=0$) shape parameters by a $0$ subscript, i.e. 
\begin{equation}
\{ \eta, \, \alpha, \, \sigma, \, \nu\, \, \zeta \}(g=0) \rightarrow \{ \eta_0, \, \alpha_0, \, \sigma_0, \, \nu_0, \, \zeta_0 \} \;. \nonumber
\end{equation}
In terms of such notation, we thus obtain from Eqs.~(\ref{eq:Q_kin}), (\ref{eq:W_grav}) and (\ref{eq:Virial2}) an expression for the soliton radius, $r_c$ in the non-interacting limit, in the  form
\begin{equation}
    r_c = \frac{2\sigma_0}{\nu_0}\frac{\hbar^2}{G M m^2} \;. \label{eq:vir_radius}
\end{equation}
 Substitution of this into the earlier reduced expression for the soliton mass Eq.~(\ref{eq:Mass}), provides a relation between $r_c$ and $\rho_0$, namely
\begin{equation}
    r_c = \left(\frac{2\sigma_0 \hbar^2 }{4\pi \nu_0 \eta_0 G m^2}\right)^{1/4}\rho_0^{-1/4}.\label{eq:vir_r_rho}
\end{equation}
Interestingly, an $r_c \propto \rho_0^{-1/4}$ relationship along with a  coefficient of proportionality were numerically obtained from fitting~\cite{Schive2014, Schive2014a} without any explicit reference to virialization. 
Our present analysis extends such numerically-based result by an analytical derivation of the coefficient of proportionality based on the empirical profile of Eq.~(\ref{eq:emp_prof}), explicitly showcasing its dependence on a range of physical parameters. Such a relationship was also previously given in~\cite{Chavanis:2020rdo}.

It is now possible to rephrase the equation for soliton mass $M$ in terms of a single parameter only, for example in terms of the peak density $\rho_0$, in the form
\begin{equation}
    M = 4\pi\eta_0\left(\frac{2\sigma_0 \hbar^2}{4\pi \nu_0 \eta_0}\right)^{3/4} \rho_0^{1/4}. \label{eq:vir_mass}
\end{equation}
Using this relation, the empirical profile in Eq.~(\ref{eq:emp_prof}) for the non-interacting limit can be reformulated as
\begin{equation}
     \rho(r)=\frac{M }{4 \eta_0 \pi r_{c}^{3}} \left(1+\lambda\left(\frac{r}{r_{c}}\right)^{2}\right)^{-8} \;. 
     \label{eq:vir_profile}
\end{equation} 

As the empirical profile provides the best approximation to the ground state of the GPPE in the non-interacting limit,
Eqs.~(\ref{eq:vir_radius}) and (\ref{eq:vir_profile}) reveal that, for any given pair of values for the soliton and boson masses $(M,m)$, one can {\em ab initio} generate a virialised soliton profile for the non-interacting case, with shape parameters taken from the empirical profile.
Such $g=0$ profile is shown in Fig.~\ref{fig:profComp},
demonstrating the good agreement with the numerical solution of the GPPE discussed in the next section. 

The profile-specific shape parameters for the three commonly used profiles are given in Table.~\ref{table:constants}, both in the absence and presence of interactions. 

\subsection{Modifications due to Interactions \label{Sec:SI}}
Next, we consider the effect of repulsive self-interactions ($g>0$) on the soliton profile. These contribute to the system's energy via Eq.~(\ref{eq:interaction_energy}).

In Sec.~\ref{sec:vir_cons} we have used the Virial theorem to obtain an expression for the core radius, $r_c$, in the non-interacting limit, Eq.~(\ref{eq:vir_radius}), expressed in terms of the non-interacting shape parameters $\sigma_0$ and $\nu_0$. Using now instead the generalized form of the virial theorem in the presence of interactions, Eq.~(\ref{eq:virial}), we obtain the more general expression in the interacting case, namely  
\begin{equation}
    r_c(g) = \frac{\sigma(\Gamma)}{\nu(\Gamma)}\frac{\hbar^2}{G M m^2}\left(1+\sqrt{1+\frac{3}{2}\frac{G m^3 M^2}{\hbar^4}\,\frac{\zeta(\Gamma) \nu(\Gamma)}{\sigma^2(\Gamma)} \,\, g}\right) \label{eq:radius_g1}.
\end{equation}
 Note that as the shape of the soliton changes with increasing value of the self-coupling $g$, \textit{all} shape parameters now become {\em a priori} unknown functions of $g$. However, we will show that such parameters become universal in terms of the dimensionless interaction strength $\Gamma = g/g_*$, so in the above equation and henceforth we denote these as~$\left\{\eta(\Gamma), \alpha(\Gamma),\sigma(\Gamma),\nu(\Gamma),\zeta(\Gamma)\right\}$.

We note here that such an equation, but with constant values for the shape parameters based on a Gaussian ansatz, was already obtained in Ref.~\cite{Chavanis2011}.
Here, we have generalized such expression to reflect the
changing soliton profile due to repulsive interactions through the $\Gamma$-dependent shape parameters. The above relation therefore describes an {\em a priori} unknown dependence of $r_c$ on $g$ which is more complex than that implied solely by the explicit appearance of $g$ on the rhs. 

The corresponding expression for the peak density can be obtained from Eq.~(\ref{eq:Mass}) as 
\begin{align}
   \rho_0(g)  = \frac{1}{4\pi\eta(\Gamma)}\, \frac{M}{r_c^3(g)} \,.
   \label{eq:rho_0_g1}
\end{align}

We can gain some immediate analytical insight for both core radius and peak density by defining corresponding limiting functions in the non-interacting and strongly-interacting limits. To do this it is useful to first define an expression for the characteristic interaction strength, $g_*$, separating these two limiting cases.
\subsubsection{Characteristic Interaction Strength}

A useful estimate of the (relative) strength of interactions can be obtained by defining a characteristic interaction strength scaling, $g_*$, as the interaction strength value at which the interaction energy equals the quantum kinetic energy~\cite{Chavanis2011}, i.e. 
\begin{equation}
   U(g=g_*)=\Theta_Q(g=g_*)
\end{equation}
\begin{equation}
   \Rightarrow  \left(\frac{\zeta (\Gamma=1)\,\, \nu (\Gamma=1)}{\sigma^2(\Gamma=1)} \right)\, g_* = \frac{10 \hbar^4}{ G M^2 m^3} \;. \nonumber \label{eq:g_*01}
\end{equation}
As the dependence of the shape parameters $\zeta$, $\nu$ and $\sigma$ on $\Gamma=g/g_*$ is {\em a priori} unknown, the above expression is in fact an implicit equation for $g_*$.
We can however simplify the procedure and obtain a well-defined explicit definition for the characteristic interaction strength $g_*$ by making the heuristic, but convenient, choice to evaluate the shape parameters using the empirical profile, Eq.~(\ref{eq:emp_prof}), in the non-interacting ($g=0$) limit.
The important benefit of our choice is that it will allow us to present results in a universal manner with $m$ and $M$ scaled out. We would expect the crossover between the non-interacting and strongly-interacting regimes to take place around\footnote{This is equivalent to the parameter $\chi$, defined in Eq. (50) in~\cite{Chavanis2011}, becoming $\chi\sim 1$.} $g\sim O(g_*)$, {\em i.e.~}$\Gamma\sim 1$.

Based on such a choice, we define $g_*$ as
\begin{equation}
   g_* \equiv \left( \frac{\sigma_0^2}{\zeta_0 \nu_0} \right) \frac{10 \hbar^4}{ G M^2 m^3} \;. \label{eq:g_*}
\end{equation}
We note that, although the actual value of $g_*$ would be slightly shifted following a different choice for the evaluation of the shape parameters, such choice would have no qualitative influence on the findings reported throughout this work. 

The effects of repulsive self interactions on the resulting soliton ground state are visible in Fig.~\ref{fig:profComp}. As one dials up the strength of the (repulsive) self interaction, the core begins to expand and the density drops, as already visible for $g = g_*$ (orange line). In this regime, neither the empirical profile nor the Thomas-Fermi profile adequately replicate the form of the density profile. However, once the interaction strength is large enough and the quantum kinetic energy is negligible in comparison to the interaction energy, the system is well modelled by the Thomas-Fermi profile, as shown there
in the case of $g = 50 g_*$.

\subsubsection{Analytical Expressions in Limiting Cases}

The above expression for $g_*$ allows us to obtain useful analytical expressions for the leading order dependence on the scaled interaction strength in the weakly and strongly interacting limits: in each such case we simplify the expression of Eq.~(\ref{eq:radius_g1}) by replacing the $\Gamma$-dependent shape parameters by the corresponding constant parameters obtained analytically in the limiting cases of zero and strong interactions (Thomas-Fermi).

In the weakly interacting case, we thus introduce
\begin{equation}
    r^0_c(g) = \frac{\sigma_0}{\nu_0}\frac{\hbar^2}{G M m^2}\left(1+\sqrt{1+\frac{3}{2}\frac{G M^2 m^3}{\hbar^4}\,\frac{\zeta_0 \nu_0}{\sigma_0^2}\,\, g }\,\, \right) \label{eq:radius_g0}\;,
\end{equation}
where subscripts 0 are used to denote corresponding shape parameter values at $g=0$.
This can now be rewritten in terms of the dimensionless interaction strength, $\Gamma$, in the more compact form 
\begin{equation}
r^0_c(\Gamma) =\frac{r_c^{NI}}{2}\left(1+\sqrt{1+15\Gamma}\right)\,,
\label{eq:gamma_radius}
\end{equation}
where $r_c^{NI}$ describes the non-interacting value given previously by Eq.~(\ref{eq:vir_radius}).

Correspondingly, in the strongly-interacting regime we introduce the notation
\begin{equation}
r^{TF}_c(\Gamma) =\frac{r^{TF}_c}{2}\left(1+\sqrt{1+15\,\frac{g_*}{g^{TF}_*}\,\Gamma}\,\,\right), \label{eq:gamma_radiusTF}
\end{equation}
where $r^{TF}_c$ corresponds to the value of $r_c$ in the limit  $g \rightarrow \infty$, and $g^{TF}_*$ is defined as in Eq.~(\ref{eq:g_*}), but based on shape parameters evaluated analytically for the Thomas-Fermi profile, i.e.
\begin{equation}
   g^{TF}_* \equiv \frac{\sigma^2_{TF}}{\zeta_{TF} \,\nu_{TF}}\,\frac{10 \hbar^4}{ G M^2 m^3} \;. \label{eq:g_*TF}
\end{equation}
The values of the shape parameters corresponding to the two limiting soliton profiles are given in Table \ref{table:constants} (columns 4 and 7 respectively for non-interacting and Thomas-Fermi limits).

Similar to the above discussion we define the limiting expressions for the interaction-strength dependence of the peak density, $\rho_0(\Gamma)$, in the non-interacting case; specifically we find  
\begin{equation}
    \rho^0_0(\Gamma) = \rho^{NI}_0\left( \frac{ 2 }{ 1+\sqrt{1+15\Gamma} }\right)^3, \label{eq:gamma_peakdens}
\end{equation}
whereas in the Thomas-Fermi limit 
\begin{equation}
    \rho^{TF}_0(\Gamma) = \rho^{TF}_0\left( \frac{ 2 }{ 1+\sqrt{1+15\,\frac{g_*}{g^{TF}_*}\,\Gamma} }\right)^3 \;. \label{eq:gamma_peakdensTF}
\end{equation}
In each above case, the shape parameters are evaluated at the corresponding limits. 
We also note here that 
\begin{equation}
    \frac{g_*}{g^{TF}_*}\simeq 0.546\,. 
\end{equation}

\subsubsection{Crossover between Weakly- and Strongly-Interacting Regimes}

The soliton size in the presence of interactions, $r_c(g)$, increases monotonically with $g$, always
lying between these two limiting curves, i.e.~$r_c^0(g) \le r_c(g) \le r_c^{TF}(g)$; this will be confirmed by numerical simulations below.
Its exact behaviour
can be directly extracted from Eq.~(\ref{eq:radius_g1}) along with a numerical determination of the shape parameters $\sigma(\Gamma)$, $\zeta(\Gamma)$ and $\nu(\Gamma)$ via the more compact relation    
\begin{equation}
    r_c(\Gamma) = \frac{\sigma(\Gamma)}{\nu(\Gamma)}\frac{\hbar^2}{G M m^2} \left(1+\sqrt{1+15\,\mathcal{C}(\Gamma)}\right)\label{eq:radius_g}
\end{equation}
where we have introduced a generalised dimensionless interaction strength parameter $\mathcal{C}(\Gamma)$ which also accounts for interaction-induced shape effects via
\begin{equation}
\mathcal{C}(\Gamma) \equiv \left[ \frac{\zeta(\Gamma)}{\zeta(0)} \,\, \frac{\nu(\Gamma)}{\nu(0)} \,\,\left(\frac{\sigma^2(\Gamma)}{\sigma^2(0)}\right)^{-1} \right]\,\,\Gamma \;.
\label{eq:Cofg}
\end{equation}
The corresponding dependence of the central density on interactions is then given by 
\begin{align}
   \rho_0(\Gamma)  = \frac{1}{4\pi\eta(\Gamma)}\, \frac{M}{r_c^3(\Gamma)} \,,
   \label{eq:rho_0_g}
\end{align}
where $\eta(\Gamma)$ can again be determined from numerically obtained soliton profiles. 
  
In the following subsection we perform a detailed numerical study which will confirm the validity of the above semi-analytical formulae. Indeed, we will see below that a numerical determination of the shape parameters, along with Eqs.~(\ref{eq:radius_g}) and (\ref{eq:rho_0_g}), describes the size and central density of a soliton of mass $M$, made up of bosons of mass $m$, very well across all values of the self-coupling $g$ from the non-interacting to the strongly interacting regimes.   
\subsection{Numerical Approach: Imaginary Time Propagation as a Tool for Ground State Profiles} 
\label{Sec:Numerical Approach}

\subsubsection{Dimensionless GPPE}
In order to numerically solve the GPPE we revert to dimensionless form by scaling all physical variables to appropriate values.
This is done in terms of a time scale 
\begin{equation}
    T=(G\rho_{\text{ref}})^{-1/2},
\end{equation}
characteristic of an object crossing a uniform density configuration of some reference density $\rho_\mathrm{ref}$ and the corresponding
energy and length scales
\begin{equation}
    E=N \hbar\sqrt{G\rho_{\mathrm{ref}}}
\end{equation}
\begin{equation}
    L=\left(\frac{\hbar^2}{m^2 G \rho_{\mathrm{ref}}}\right)^{1/4}
\end{equation}
for some particle number $N=M/m$.
Setting $t=T t'$ and $\mathbf{r}=L \mathbf{r}'$, the resulting dimensionless GPPE takes the form
\begin{equation}
\displaystyle i\frac{\partial \psi\prime}{\partial t^\prime} = -\frac{ \nabla^{\prime2}}{2} \psi^\prime + V^\prime\psi^\prime + g^\prime|\psi^\prime|^2 \psi^\prime
\end{equation}
and
\begin{equation}
\nabla^{\prime2} V^\prime = 4\pi \left(\frac{\rho_\text{sys}}{\rho_{\text{ref}}}\right)(\rho^\prime- \bar{\rho}^{\prime}),
\end{equation}
where the interaction strength is written as $g = g^{\prime} \hbar\sqrt{G\rho_{\mathrm{ref}}}  /\rho_{\mathrm{sys}} $, the gravitational potential as $V=V' \hbar \sqrt{G}\rho_{\mathrm{sys}}/m\sqrt{\rho_{\mathrm{ref}}}$ and the wavefunction has been  scaled to the mean density $\rho_{\mathrm{sys}}$ within the computational box through the dimensionless form $\psi^\prime=\psi/\sqrt{\rho_\mathrm{sys}}$.
We choose the mean reference density of the system as
$\rho_{\mathrm{ref}} = 1.5 \times 10^3 M_\odot/\mathrm{kpc}^3$ which is commensurate with the current cosmic density.  We will henceforth drop all primes for the sake of simplicity.

As the true ground state solution to the GPPE (which is what we are seeking numerically) is a spherically symmetric density distribution, we further simplify the problem to be solved numerically to a spherically symmetric form (thus disallowing any angular dependence/asymmetries to creep up also in subsequent dynamical analysis). A further significant benefit in doing so is that it makes the required numerical resolution of the soliton feasible, without unnecessarily strong demands on computational memory and time.

Using the further substitution $\psi = \phi/r$, we obtain the radial (one-dimensional) GPE in the form
\begin{equation}
    i\frac{\partial \phi}{\partial t}=\left(-\frac{1}{2}\frac{\partial^2}{\partial r^2} + V +g\left|\frac{\phi}{r}\right|^2\right)\phi \label{eq:GPE_spc}
\end{equation}
which we solve using the implicit midpoint method. 
The Poisson equation to which the GPE is coupled to now takes the simpler form
\begin{equation}
\frac{\partial^2 \mathcal{V}}{\partial r^2} = 4\pi r\rho
\end{equation}
where we have made the substitution $V = \mathcal{V}/r$ and dropped the  $\bar{\rho}$ term which is no longer necessary due to the fact that we are using a spherically symmetric regime with boundary conditions that go to zero. (Note that subtraction of the mean density would have been necessary if we were to include periodic boundary conditions.) 

The computational speed up acquired by translating to a spherically symmetric system and using this method is significant and allows for the study of higher mass solitons with larger resolution, where 3D approaches may suffer from a lack of computational power or simply from a lack of memory in being able to store large 3D arrays. 

\subsubsection{Imaginary Time Propagation \& Numerical Ground State}

Early studies of the ground state soliton employed a shooting method to obtain the solution of the time-independent GPPE equations which decays at infinity \cite{Chavanis2011_2,Guzman:2006yc} or utilized a ``shedding'' of the excited modes which escaped the computational box imposed by an absorbing boundary \cite{Guzman:2006yc}. Here, we take a different approach to generate the ground state solution of the system, employing a method widely used in various guises in the study of cold atomic condensates~\cite{Minguzzi-Imaginary-Time, PhysRevA.51.4704, PhysRevLett.82.1616}  by propagating our system in imaginary time (but see also \cite{Chavanis2020-2, Dmitriev2021} in the FDM context). By setting $dt \rightarrow - i dt$, we reformulate the system in such a way that a `forward' propagation in {\em imaginary} time results in all higher energy eigenstates decaying exponentially at a rate proportional to the energy of each eigenstate. As a result, the ground state decays comparatively more slowly than all other (excited) states. By additionally enforcing a renormalization of the particle number within the system at the end of each $dt$ propagation, we ensure that the norm of $\phi$ remains constant through such a procedure. Such combination of imaginary time propagation and renormalization gradually extracts energy from the system and suppresses all occupied excited eigenstates present in the initial configuration, while keeping the total number of particles fixed. As a direct consequence of this, it directly enforces a gradual transition to the ground state without the need for resorting to either a shooting method or a removal of modes via absorbing boundary conditions.

In our calculations we assume that the ground (virialized)  state is reached when the total energy of the system has converged 
to within one part in  $10^{8}$. To ensure the size of our numerical grid has no influence on our findings, we also monitor the value of the density at $r = \Delta r$, the inner-most point of the computational sphere, and also the ratio $\rho(\Delta r)/\rho(L)$, where $L$ is the edge of the computational sphere. 
By imposing a large value for this ratio (here $\rho(\Delta r)/\rho(L) > 10^{40}$), we ensure that the computational sphere is large enough in size to adequately contain the soliton without the boundaries interfering with its shape and dynamics. 

Moreover, in our numerical calculations, we need to balance the need for good accuracy of the soliton core (i.e.~high spatial grid resolution) -- which controls the extent to which the virial theorem is numerically satisfied -- and the computational time required (which increases with increasing number of grid points). In practice we found that the description can be sufficiently accurate upon including at least 5 grid points between $r = 0$ and $r = r_c$, although a higher number of grid points will further enhance numerical accuracy. We find that the virial condition (\ref{eq:virial full}) is satisfied by our solutions on the order of 1\%.

Once the ground state solitons have been obtained from the imaginary time propagation, they are then propagated in real time to confirm that they are indeed solutions close to the ground state. 
When evolved in real time, any deviations from the true solitonic ground state will result in the creation of an ``atmosphere'' of bosons around the soliton - a miniature version of a dark matter halo, 
formed by excited bosons ejected from the soliton\footnote{Unlike the simulations presented in the early work of \cite{Guzman:2006yc} which shed modes above the ground state soliton via an absorbing boundary condition, a process termed there gravitational cooling, we here effectively form mini-halos which remain gravitationally bound, fitting comfortably within our computational box.}. 
somewhat reducing its mass $M$. To estimate this deviation from the ground state we calculate the total mass residing in the temporally averaged soliton across the entire dynamic simulation. This is achieved by numerically determining the Penrose-Onsager mode as in~\cite{Liu2023}, which defines the true ground state. We thus verify that only a very small mass leakage from the soliton occurs: in all our simulations losses to atmospheres around the core are at a level of at most $10^{-5}$ of the initial ground state solution's mass $M$.

Shape parameter values are extracted from simulations by performing the relevant numerical energy integrals on the generated ground state profile and then using the peak density and core radius of the numerical soliton to calculate them. Although all shape parameters should be uniquely identified in the case of $g = 0$, numerically we find a slight variance of up to few $\%$ in their values, as seen in the fifth column of Table~\ref{table:constants}.
Such variance accounts for a slight change due to a different spatial resolution and computational box size.
However, in all simulated non-interacting cases, there is a clearly identified value for each shape parameter, with such values being in accordance with our analytical values calculated from the empirical formula (fourth column of Table~\ref{table:constants}). The numerical average for each non-interacting shape parameter value is plotted on the left of each subplot in (subsequent) Fig.~\ref{fig:allCoeffs} as a filled green point at $\Gamma = 0$. Error bars are not visible in most cases, due to the small standard deviation in results. 

\subsubsection{Numerical Results for Variable Self-interaction Strength}

\begin{figure}
    \centering
    \includegraphics[width=0.9\linewidth,keepaspectratio]{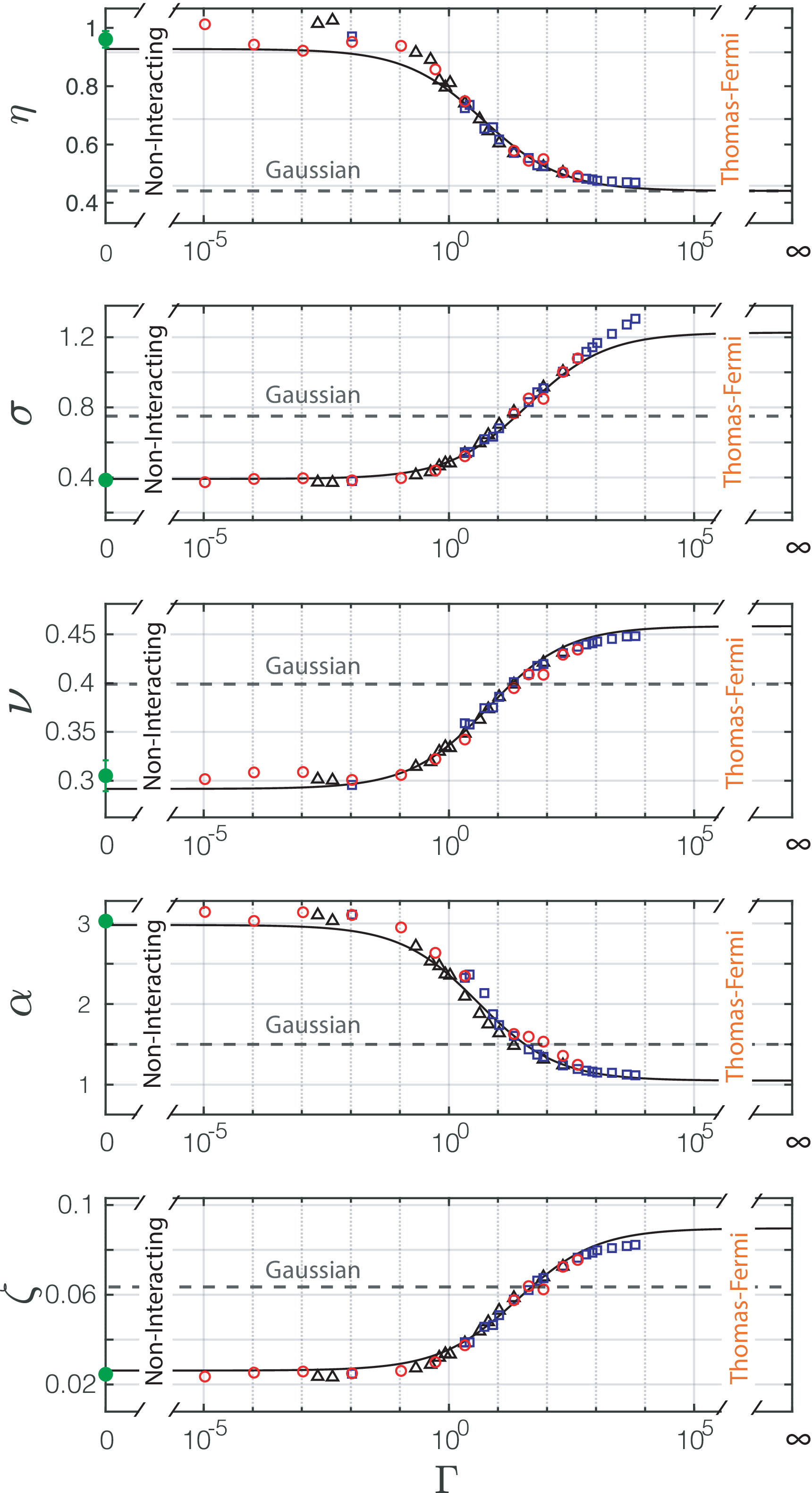}
    \caption{
 Dependence of the value of shape parameters on interaction strength, scaled to the characteristic value $g_*$ defined by Eq.~(\ref{eq:g_*}).
Such dependence is reproduced by a two-parameter fit based on Eq.~(\ref{eq:shape_fit}) which has as limiting cases the non-interacting limit at $\Gamma = 0$ (left), and the Thomas-Fermi limit at $\Gamma \rightarrow \infty$ (right), whose exact values for each parameter are given in Table \ref{table:constants}. 
The corresponding predictions of Ref.~\cite{Chavanis2011} based on a Gaussian approximation (reported in Column 3 of Table~\ref{table:constants}) are shown here by the horizontal dashed line. Filled green points at the $g=0$ line 
represent the averaged values of the corresponding shape parameters from numerical simulations (see Table~\ref{table:constants}, fifth column).
    }
    \label{fig:allCoeffs}
\end{figure}

Ground states were generated for a range of $g$ values, for several different configurations of $m$ and $M$. 
Shape parameter values were extracted from these numerical ground states. 
Such behaviour is shown for all 5 shape parameters ($\eta$, $\sigma$, $\nu$, $\alpha$ and $\zeta$) in Fig.~\ref{fig:allCoeffs}. 
As evident, all such parameters follow a clear monotonic transition from the non-interacting limiting case (leftmost part, green points) to the Thomas-Fermi case (rightmost part), with some of them decreasing ($\eta$, $\alpha$) and others increasing ($\sigma$, $\nu$, $\zeta$) with increasing (scaled) interaction strength $\Gamma$.

Interestingly, we find that the dependence of all such shape parameters on $\Gamma$ can be excellently fit by a function of the form
\begin{equation}
    u(\Gamma) = \frac{u_0+u_{TF}}{2}+\frac{u_0 - u_{TF}}{2} \left[ 1 - \tanh\left( \frac{\log_{10}\left(\Gamma\right)-a}{b}  \right) \right] \; , \label{eq:shape_fit}
\end{equation}
where $u$ denotes any of these shape parameters, subscripts 0 and $TF$
denote the corresponding analytical values in each regime, and $a$ and $b$ are fitting constants.
The specific chosen values of these fitting constants for each shape parameter are given in Table~\ref{table:shape_fit}, and vary slightly for each shape parameter. Nevertheless, in all cases, both $a$ and $b$ are of order unity.

\begin{table}[]
\begin{tabular}{|l|l|l|l|l|l|}
\hline
  & $\eta$ & $\sigma$ & $\nu$ & $\alpha$ & $\zeta$ \\ \hline
$a$ & 0.4   & 1.15    & 0.5  & 0.3     & 1.1    \\ \hline
$b$ & 1.5    & 1.5      & 1.7   & 1.5      & 1.5     \\ \hline
\end{tabular}
\caption{The fitting constants for Eq.~(\ref{eq:shape_fit}).}
\label{table:shape_fit}
\end{table}

An important comment needs to be made regarding the shape parameter $\sigma$, which originates from the quantum pressure. Given that the Thomas-Fermi regime is defined by a negligible quantum pressure, the analytical value computed from the quantum kinetic energy integral for the Thomas-Fermi profile loses meaning. 
Moreover, the deviation of the numerical values  for $\sigma$ from the fit in Fig.~\ref{fig:allCoeffs} for large values of $\Gamma > 10^3$, can be attributed to the much steeper decrease of the Thomas-Fermi profile towards a zero value (in stark contrast to the smoother decrease of the numerically generated profiles).

Next, we comment on the validity of the Gaussian approximation for different values of $g$. As is evident from Fig.~\ref{fig:allCoeffs}, the Gaussian values taken from Ref.~\cite{Chavanis2011} always fall within  our obtained numerical range: in fact, such points typically lie close to the midpoint between our presented analytical limiting values (except for $\eta$ where the Gaussian value effectively corresponds to the Thomas-Fermi limit), and, as such, are  good approximate values in the crossover regime. Note however that although such values are never off by more than a factor of $3$, such analytically convenient choice of values cannot capture the true variable nature of the shape parameters. 

\begin{figure*}
\centering 
\includegraphics[width=1\linewidth, keepaspectratio]{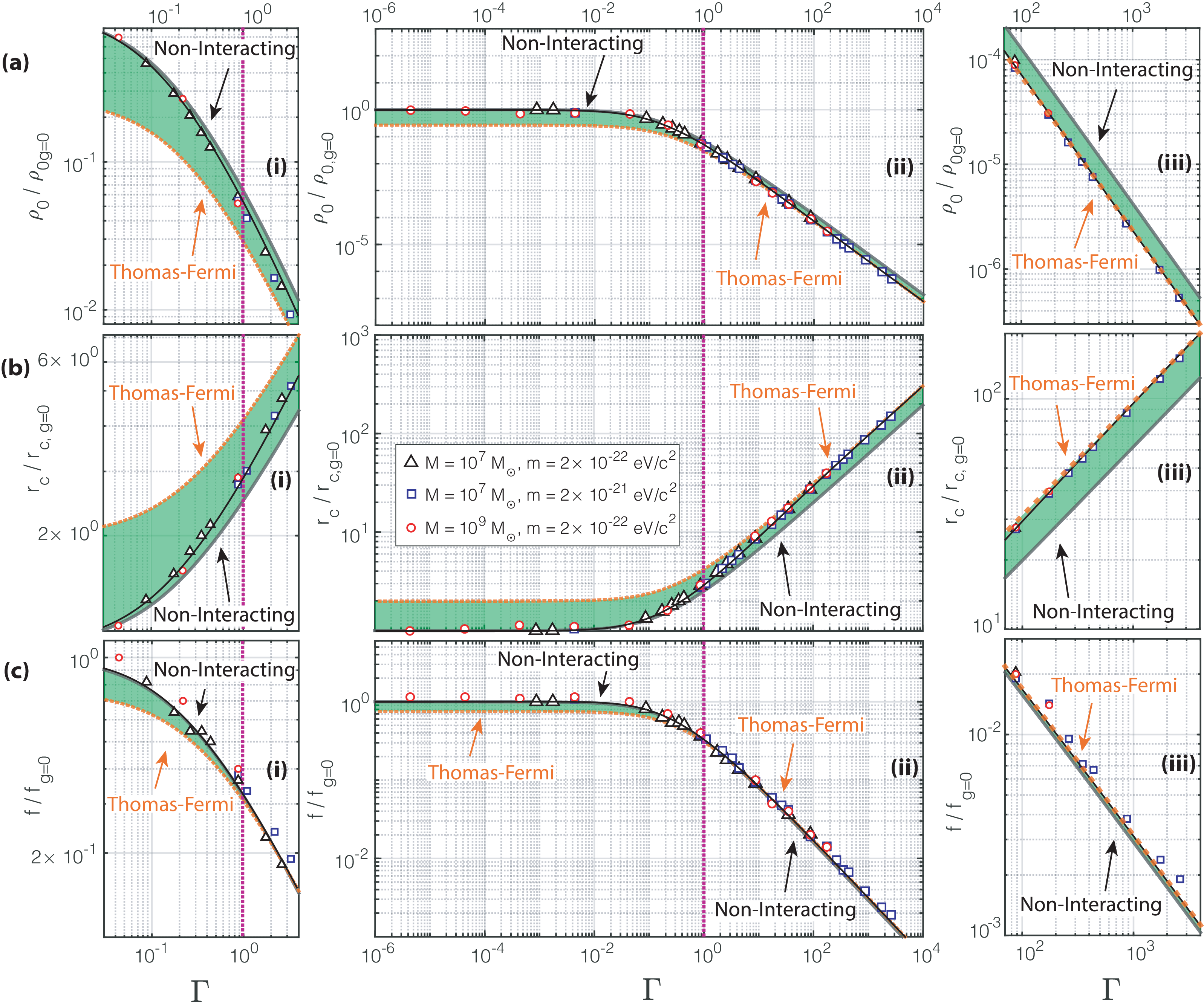}
 \caption{ Universal dependence of key static and dynamical soliton properties on scaled dimensionless interaction strength. Shown are [from top to bottom] the cases for (a) peak soliton core density, (b) core radius, and (c) core oscillation frequency, with such quantities scaled to their corresponding non-interacting simulated values, and all self-interactions scaled to the characteristic value $g_*$ from Eq.~(\ref{eq:g_*}).
All plots show numerically simulated data for 3 different ($M$, $m$) combinations and a continuous solid black line obtained from our analytical equations (\ref{eq:radius_g-ratio}), (\ref{eq:rho_g-ratio}), (\ref{eq:gamma_freq}) using as input the $\Gamma$-dependent shape factors shown in Fig.~(\ref{fig:allCoeffs}). 
Panels (ii) [middle] reveal the entire crossover, with (i) left and (iii) right panels respectively highlighting the weakly-interacting and strongly-interacting limits.
The two limiting boundaries of the 
green channels highlight the range of accessible predicted values, bounded from above and below by the respective characteristic universal curves: specifically, 
the non-interacting grey line is constructed from Eqs.~(\ref{eq:gamma_peakdens}), (\ref{eq:gamma_radius}), (\ref{eq:gamma_freq})), with $g_*$ [Eq.~(\ref{eq:g_*})] computed using the non-interacting shape parameters; moreover, the dashed black Thomas-Fermi line arises from Eqs.~(\ref{eq:gamma_peakdens}), (\ref{eq:gamma_radius}), (\ref{eq:gamma_freq}) with $g_*$ [Eq.~(\ref{eq:g_*})] calculated instead using the Thomas-Fermi shape parameters.
The vertical purple line highlights the characteristic interaction strength value $g=g_*$ (corresponding to $\Gamma=1$). Note that a cross-over between the limiting non-interacting and Thomas-Fermi lines is in fact present in the case of the oscillation frequency [(c)(ii)], despite this being largely obscured by the very narrow accessible channel. 
We highlight that the excellent fit of our semi-analytical predictions (solid back lines) through all our numerical data demonstrates the importance of the correct incorporation of the universal variation of shape parameters on scaled interaction strength. Remarkably, such an approach accurately predicts the peak density, radius and frequency, even in the transition region between non-interacting and Thomas-Fermi limits.
 }
\label{fig:interacting_data}
\end{figure*}

Having fully quantified the universal dependence of the shape parameters on the dimensionless interaction strength $\Gamma=g/g_*$, we can now discuss the corresponding, and observationally-relevant, dependence of the key soliton profile parameters of peak density and soliton radius on interactions. To present this in the most general manner, we consider the dependence of the rescaled central soliton density $\rho_0(\Gamma)/\rho_0(0)$ and core radius $r_c(\Gamma)/r_c(0)$ as a function of the rescaled self-interaction strength $\Gamma=g/g_*$.
The results of our numerical simulations
are respectively shown in Fig.~\ref{fig:interacting_data}(a)-(b). 

Superimposing numerical data points for 3 different total mass $M$ and soliton mass $m$ combinations (in the range $2\times 10^{-22} \, \mathrm{eV/c^2} < m < 2 \times 10^{-21} \, \mathrm{eV/c^2}$ and $10^7 \, \mathrm{M}_\odot < M < 10^9 \, \mathrm{M}_\odot$), shown by different symbols/colours, the dependence of scaled peak density and soliton core on $\Gamma$ is shown over the entire probed range of $\Gamma \in [4\times 10^{-6}, \, 3\times 10^3]$. 

Analytical expressions for these curves can also be obtained as follows:
The scaled soliton width dependence on $\Gamma$ can be obtained through Eq.~(\ref{eq:radius_g}) as
\begin{equation}
\frac{r_c(\Gamma)}{r_c(0)} = \left( \frac{\sigma(\Gamma)}{\sigma_0} \right) \left(\frac{\nu(\Gamma)}{\nu_0}\right)^{-1}
\left(\frac{1+\sqrt{1+15\,\mathcal{C}(\Gamma)}}{2}\right) \;. \label{eq:radius_g-ratio}
\end{equation}
The corresponding dependence of the peak density can be obtained from Eq.~(\ref{eq:rho_0_g}) via
\begin{equation}
\frac{\rho_0(\Gamma)}{\rho_0(0)} = \left( \frac{\eta(\Gamma)}{\eta_0} \right)^{-1} \left(\frac{r_c(\Gamma)}{r_c(0)} \right)^{-3} \;. \label{eq:rho_g-ratio}
\end{equation}
Inputting our numerically-evaluated universal shape parameter dependence on $\Gamma$ based on Eq.~(\ref{eq:shape_fit}) for $\sigma(\Gamma)$, $\nu(\Gamma)$, $\zeta(\Gamma)$, and $\eta(\Gamma)$ into the above equations yields the solid black lines in Fig.~\ref{fig:interacting_data}(a)-(b): such curves are clearly found to lie directly on top of all our numerical points for different $(M,\,m)$ combinations. To clarify such observations, in each of the two cases we also plot the corresponding fully-analytical weakly-interacting ($r_c^0(\Gamma)$, $\rho_0^0(\Gamma)$) [solid grey lines] and strongly-interacting (Thomas-Fermi) limits ($r_c^{TF}(\Gamma)$, $\rho_0^{TF}(\Gamma)$) [dotted orange lines], defined respectively by Eqs.~(\ref{eq:gamma_radius}), (\ref{eq:gamma_peakdens}) and (\ref{eq:gamma_radiusTF}), (\ref{eq:gamma_peakdensTF}). These curves delimit the region where all our data points lie and which we indicate by green shading.

In both cases we clearly see a transition from the non-interacting behaviour for small $\Gamma$ to the Thomas-Fermi behaviour for large $\Gamma$, with such transition indeed occurring around $\Gamma=1$, i.e.~near the characteristic interaction strength $g_*$ of Eq.~(\ref{eq:g_*}); such value has been highlighted by a vertical dashed red line for easier visualization.
 To make such transitions between the two limiting cases more transparent, we further highlight the weakly- and strongly-interacting limits in subplots (i) and (iii), respectively found to the left and right of the main plot.

We have thus fully characterized our soliton ansatz of Eq.~(\ref{eq:general_ansatz}) in terms of a universal dependence on the (scaled) interaction strength, with such behaviour obtained semi-analytically (by a combination of analytical expressions combined with the universal dependence of the shape parameters on $\Gamma$) and confirmed by full numerical simulations.

\section{Soliton Oscillations \label{Sec:soliton_oscillations}} 

The simplest (lowest-energy) excitation of the soliton core comes in the form of a radial oscillation. The rate of such an excitation
depends on the total mass of the soliton, the mass of the constituent boson and—in the interacting case—the interaction strength \cite{Chavanis2011}. Soliton core oscillations could in fact lead us to signals of FDM's existence in astronomical observations, as well as allow us to place constraints on the range of possible boson masses \cite{Marsh2019, Chiang2021-2}, and they may play a part in gravitational heating of galaxies. It is also possible that core oscillations may drive dynamical behaviour in the entire halo structure due to interference \cite{Chowdhury2023}.

 In this section we examine the dependence of the frequency of such small radial oscillations on the self-coupling strength parameter in terms of the dimensionless ratio $\Gamma$. This will allow us to overlay data from a variety of solitons on a universal curve.     

\subsection{Isotropic oscillation}

A proper study of the soliton's oscillations would involve a thorough analysis of its normal modes, an investigation that we leave for an upcoming publication. Here, following \cite{Chavanis2011, Chavanis:2020rdo}, we will employ a perturbative method for studying the radial oscillatory dynamics of the soliton, making the further assumptions that: (i) small oscillations can be described by making $r_c$, the parameter that sets the scale of the solitonic profile Eq.~(\ref{eq:general_ansatz}), time-dependent, i.e.~$r_c\rightarrow r_c(t)$, and (ii) the resulting velocity field takes the isotropic form \begin{equation}\label{eq:unif_exp}
 \mathbf{u} = f(t) \, \mathbf{r}  \,. 
\end{equation}
As we now show, these two assumptions can satisfy the continuity equation, Eq.~(\ref{eq:continuity}), for a specific form of the function $f(t)$. 

Although the form of the soliton profile $\rho(r)$ is not known for $g \neq 0$, we may make use of our ansatz $\rho(r)=\rho_0 \varphi(r/r_c,\,\Gamma)$, 
further assuming that the profile becomes time dependent only via $r_c\rightarrow r_c(t)$. We thus obtain
\begin{align}
\frac{\partial \rho}{\partial t} =-\frac{\partial \ln r_{c}}{\partial t} \left(\frac{\rho_0}{r_{c}}\varphi'r + 3 \rho\right)
\end{align}
and     
\begin{align}
    \nabla \cdot (\rho\mathbf{u}) = f(t) \left(\frac{\rho_0}{r_{c}}\varphi'r + 3 \rho\right).
\end{align}
where $\varphi'$ denotes a derivative w.r.t. the function's spatial argument. Clearly, the continuity equation is fulfilled if 
\begin{equation}
    f(t) = \frac{1}{r_c} \frac{\partial r_c}{\partial t}
\end{equation}
and therefore describing the oscillation by assuming that the whole profile evolves with time via $r_c(t)$ and with the velocity profile 
\begin{equation}\label{eq:unif_exp}
 \mathbf{u} = \frac{1}{r_c} \frac{\partial r_c}{\partial t} \mathbf{r}  \,, 
\end{equation}
is consistent witn mass conservation. This allows us to calculate the classical kinetic energy, $\Theta_C$, as
\begin{align}
    \Theta_C &= \frac12 \int \rho |\mathbf{u}|^2 d^3r,\\
    &= \frac12 \left(\frac{d r_c}{dt}\frac{1}{r_c}\right)^2\left[4\pi \int_0^\infty \rho r^4 dr\right],\\
    &=\frac12 \alpha M\left(\frac{d r_c}{dt}\right)^2. \label{eq:Theta_c}
\end{align}
Therefore, solitons can exhibit oscillations which physically manifest as a uniform expansion and contraction of the soliton profile. The soliton mass $M$ is constant in time and therefore the peak density must also experience oscillations as $\rho_0(t) \propto r^{-3}_c(t)$ - see Eq.~(\ref{eq:Mass}).

\subsection{Analytical Oscillation Frequencies}

The total energy of the interacting system can be written as 
\begin{align}
    E_{\text{tot}} &= \Theta_C+\Theta_Q+W+U\\
    &= \frac12 \alpha M\left(\frac{d r_c}{dt}\right)^2 + \sigma \frac{\hbar^{2}}{m^{2}} \frac{M}{r_{c}^{2}} - \nu\frac{GM^2}{r_c} + \zeta\frac{g M^2}{2m r_c^3} \\
    &= \frac12 \alpha M\left(\frac{d r_c}{dt}\right)^2 + V(r_c)\;,
\end{align}
which can be thought of as an integral of the motion for the dynamical equation
\begin{equation}
    \alpha M \frac{d^2 r_c}{dt^2}= - \frac{dV}{dr_c}. \label{td_virial}
\end{equation}
Hence, the fundamental, breathing oscillation mode of the soliton can be described by the one-dimensional motion of a non-relativistic, Newtonian particle moving in the potential $V$~\cite{Chavanis2011} defined above.

A static solution corresponds to a $g$-dependent equilibrium radius, $r^*_c$, satisfying
\begin{equation}
    \frac{dV}{dr_c}\Bigg|_{r^*_c} = 0\,. 
\end{equation}
By performing a small perturbation about this equilibrium radius $r_c(t) = r^*_c + \varepsilon(t)$ in (\ref{td_virial}), we find
\begin{equation}
\alpha M \ddot{\varepsilon} (t)+\left( 6\sigma \frac{\hbar^2}{m^2}\frac{M}{r_c^4}-2\nu\frac{G M^2}{r_c^3}+12\zeta\frac{M^2g}{2m r_c^5}\right)\varepsilon (t) = 0,
\end{equation}
where, to avoid notational clutter, we henceforth drop the $*$ superscript. At this point we remind the reader that, as $V(r_c)$ depends on $g$ both explicitly and implicitly via the shape parameters $\sigma=\sigma(\Gamma)$, $\nu=\nu(\Gamma)$ and $\zeta=\zeta(\Gamma)$, such equilibrium value depends on $g$: in fact it corresponds to the $g$-dependent equilibrium value given by Eq.~(\ref{eq:radius_g1}). 
We thus infer~\cite{Chavanis2011} a frequency 
\begin{equation} 
    f = \frac{1}{2\pi}\sqrt{\frac{6\Theta_Q+2W+12U}{I}} \;, \label{eq:freq_unvirialised}
\end{equation}
where we remind the reader that $I=\alpha M r_c^2$ is the moment of inertia.
Accounting for the interacting virial condition [Eq.~(\ref{eq:virial full})] and using the expressions for the shape parameters we can rewrite this as
\begin{equation}
f(g)  = \sqrt{\frac{\sigma(\Gamma)}{2\,\alpha(\Gamma) \pi^2}\,\frac{\hbar^2}{m^2 \,\, r_c^{4}(g)}\, \frac{1+\left(\frac{3\,\zeta(\Gamma) }{2\,\nu(\Gamma)}\,\frac{1}{ G m \,\, r_c^{2}(g)}\right)\,g}{1-\left(\frac{3\,\zeta(\Gamma) }{2\,\nu(\Gamma)}\,\frac{1}{ G m \, r_c^{2}(g)}\right)\,g}}\;. \label{eq:frequency_g}
\end{equation}
Note that $r_c$ is now a function of the self-coupling $g$.
Such an expression, but with {\em constant} shape factors, has been previously analytically derived in Ref.~\cite{Chavanis2011,Chavanis:2020rdo}.
Here we generalize this to include the numerical differences hidden within the the dpendence of the shape parameters on $\Gamma$.

In the $g\rightarrow 0$ limit and by making use of Eq.~(\ref{eq:vir_r_rho}),  we can write the oscillation frequency 
in a form which is solely dependent on the peak density of the soliton,
\begin{equation}
    f(0) = \left(\frac{G\nu_0\eta_0}{\alpha_0 \pi} \right)^{1/2}\,\rho_0^{1/2}, \label{eq:freq_rho}
\end{equation}
with the shape parameters $\nu_0, \, \eta_0$ and $\alpha_0$ for the $\Gamma=0$ empirical profile, Eq.~(\ref{eq:emp_prof}), given in Table I.  
This form agrees with the $f \propto \rho_0^{1/2}$ relationship expressed in previous literature~\cite{Veltmaat2018, Guzman2019, Schive2020, Veltmaat2020-Baryons, Chiang2021-2, Chowdhury2021}. Specifically, we find 
\begin{equation}\label{eq:freq_r}
    f(0) = 11.4 \left(\frac{\rho_0}{10^9 M_\odot \mathrm{kpc}^{-3}}\right)^{1/2}\mathrm{Gyr}^{-1}\,,
\end{equation}
while in e.g. \cite{Veltmaat2018} the coefficient is $10.94$, a value obtained from analysing oscillations of the central soliton in a simulated core-halo system. Very similar values are quoted in the literature~\cite{Guzman2019, Schive2020, Veltmaat2020-Baryons, Chiang2021-2, Chowdhury2021}. The difference is very small and could be due to the fact that values quoted in the literature are mostly extracted from solitons embedded in halos and not in isolation which may not be oscillating in their fundamental frequency only.  

It is interesting to note that by looking at Eq.~(\ref{eq:vir_radius}) and Eq.~(\ref{eq:freq_r}) we can compare our results to those from \cite{Chavanis2011}, finding that
\begin{eqnarray}
    &r_{\mathrm{emp}} &= 0.7149 
 r_\mathrm{Gauss}, \\
    &f_\mathrm{emp} &= 1.0021 
 f_\mathrm{Gauss}.
\end{eqnarray}
As a result, and perhaps somewhat unexpectedly,
 although the radius parameters between the Gaussian profile and the empirical profile (\ref{eq:emp_prof}) are clearly distinct, the balancing of these and the various shape parameter values in these limiting cases appearing in Table \ref{table:constants} 
 result in almost perfect cancellation, thus leading to practically the same predictions for the {\em frequency}. Therefore, our extended present work confirms that the Gaussian ansatz approach can be considered a remarkably robust approach for analysing the soliton's oscillation frequency.

\subsection{Numerical Results for Non-Interacting Solitons}
\label{sec:Simulation non-interacting solitons}
\begin{figure*}
    \includegraphics[width=1\linewidth,keepaspectratio]{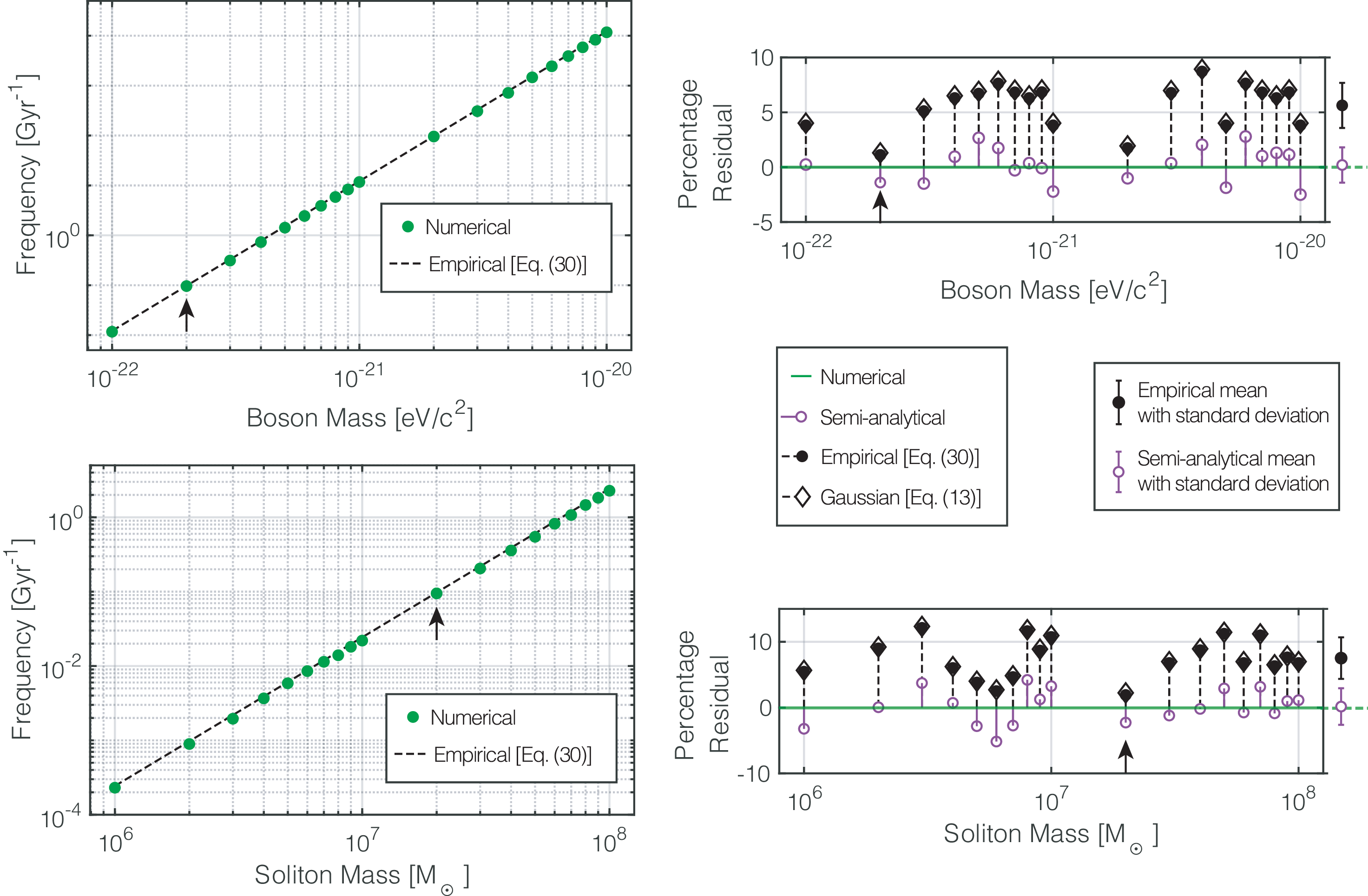}
    \caption{\textit{Left:} (Top) The oscillation frequency for varying boson mass for a total soliton of mass $M_{\mathrm{soliton}} = 2\times10^7$ M$_\odot$. (Bottom) The various results for oscillation frequency when the soliton mass is varied from a range of $10^6$ to $10^8$ M$_\odot$. The black arrow indicates a mutual point which is present in both simulation runs. The dashed line is plotted according to Eq.~(\ref{eq:freq_mM}), in which the shape parameters are calculated from the relevant energy integrals of the empirical profile, Eq.~(\ref{eq:vir_profile}). 
    \textit{Right:} A residual plot of the oscillation frequencies, calculated according to $\left[(\textrm{simulation} - \textrm{analytics})/\textrm{simulation}\right]$, for the case of (Top) varied boson mass and constant soliton mass, and (Bottom) varied soliton mass and constant boson mass. The zero line represents the simulation data. The filled black data points correspond to the residual between the simulation data and Eq.~(\ref{eq:freq_mM}), with shape parameters from the energy integrals using the empirical profile. The black diamonds correspond to the same, but instead making use of the shape parameters from the Gaussian. The hollow purple circles use a numerical extraction of the shape parameters from each ground state solution to calculate the frequency. In either case, residuals are grounded on simulation data. To the right of each residual plot is a data point which represents the mean residual value for the case of frequency calculations from the empirical profile, Eq.~(\ref{eq:vir_profile}), and the corresponding variance in the form of error bars. As can be seen from the two right sub-plots and discussed in the text, using the empirical or Gaussian profiles gives essentially identical results for the oscillation frequencies which are less accurate than those determined by using the shape parameters obtained from the numerical soliton profiles.
   }
    \label{fig:m_M_Final}
\end{figure*}

To numerically study the lowest-energy soliton oscillations we must perturb the obtained ground states. This can be very easily engineered in our numerical simulations using the following trick: instead of running our imaginary time propagation until full system equilibration 
has been achieved, we terminate such process somewhat earlier to allow for the otherwise almost perfect ground state solution to be left with a `natural' built-in perturbation. Subsequently propagating such perturbation in the (real) time domain we can extract the frequency of such an oscillatory mode by performing a Fourier analysis of the dynamics of the peak density value.

According to Eq.~(\ref{eq:vir_radius}) the radius, and therefore the peak density, are both functions of boson mass. We can therefore rewrite Eq.~(\ref{eq:freq_rho}) in terms of the boson mass and soliton mass,
\begin{equation}
    f(0) = \frac{1}{2\pi} \left( \frac{\nu_0^4}{8 \alpha_0 \sigma_0} \frac{G^4 M^4 m^6}{\hbar^6}\right)^{1/2}. \label{eq:freq_mM}
\end{equation}
In order to compare results of the empirical and Gaussian profiles to our numerics, 
we evaluate the above frequency formula, Eq.~(\ref{eq:freq_mM}), using the shape parameters obtained from the respective profiles.
A detailed comparison between simulation data and the prediction from Eq.~(\ref{eq:freq_mM}) is shown in Fig.~\ref{fig:m_M_Final} as a function of both (i) changing boson mass (within the range $10^{-22} \mathrm{eV/c}^2\le m \le 10^{-20} \mathrm{eV/c}^2$) [top plots] and (ii) changing soliton mass (within the range $10^{6} M_\odot \le M \le 10^{8} M_\odot$) [bottom plots]. This confirms the very good overall validity of Eq.~(\ref{eq:freq_mM})
for the empirical (and Gaussian) analysis with our numerical data [Fig.~\ref{fig:m_M_Final}(a)].
To better understand how these compare, and what the subtle differences between empirical and Gaussian results (concealed in Fig.~\ref{fig:m_M_Final}(a)) may be, Fig.~\ref{fig:m_M_Final}(b) plots their scaled residual differences in each case.

This reveals that, despite a very good overall agreement with our numerical simulation data, such results for both empirical (filled black circles) and Gaussian (hollow diamonds) consistently overshoot the numerically-obtained frequency (green line).
Motivated by our preceding detailed shape-parameter analysis, we thus proceed to
calculate the shape parameters individually from each numerically generated ground state, and use these -- rather than the shape parameters from the empirical profile -- in Eq.~(\ref{eq:freq_mM}): as expected, this semi-analytical procedure (open purple circles) yields a much better agreement with the oscillation frequency extracted from simulations. It is therefore clear that the calculation of the oscillation frequency of a soliton is in fact also sensitive (on the $10\%$ level) to the shape parameters and therefore the shape of the profile. 
\subsection{Results for Interacting Solitons}

The analytical prediction (\ref{eq:frequency_g}) for the oscillation frequency in the case of $g\neq 0$ can be written in terms of the dimensionless parameter $\Gamma$ as
\begin{eqnarray}
    \frac{f(\Gamma)}{f(0)} &= &\left( \frac{ 2 }{ 1+\sqrt{1+15\,\mathcal{C}(\Gamma)} }\right)^2  \nonumber \\
    &\times & \sqrt{1+\frac{15\,\mathcal{C}(\Gamma)}{1+\sqrt{1+15\,\mathcal{C}(\Gamma)}}}
    \label{eq:gamma_freq}
\end{eqnarray}
where we remind the reader that $\mathcal{C}(\Gamma) \equiv \frac{\zeta(\Gamma)\nu(\Gamma)}{\zeta(0)\nu(0)}\frac{\sigma^2(0)}{\sigma^2(\Gamma)}\,\Gamma$ - see Eq.~(\ref{eq:Cofg}). A comparison of the above formula, with the shape parameters obtained from numerical soliton profiles, to numerical simulations of oscillating solitons are   shown in fig.~Fig~\ref{fig:interacting_data}(c).

Once again, as in the analysis of $\rho_0$ and $r_c$ of the previous section, we can obtain the limiting values for $f$ in the cases $\Gamma=0$ and $\Gamma \rightarrow \infty$, by evaluating all shape factors $\{\zeta(\Gamma),\,\nu(\Gamma),\,\sigma(\Gamma)\}$ appearing in $\mathcal{C}(\Gamma)$ in terms of their non-interacting limits  $\{\zeta(0),\,\nu(0),\,\sigma(0)\}$, or their strongly-interacting Thomas-Fermi limits $\{\zeta^{TF},\,\nu^{TF},\,\sigma^{TF}\}$.
Such process gives us the corresponding non-interacting and Thomas-Fermi channels for the frequencies, respectively shown by the grey and dashed orange lines in Fig~\ref{fig:interacting_data}(c).
Interestingly, in our numerical results the two channel limits experience a cross-over around $\Gamma \approx 1$, as can be seen in Fig~\ref{fig:interacting_data}(c)(i)-(iii) which show that the limit at the top of the channel is the non-interacting one for $\Gamma \lessapprox 1$ [Fig~\ref{fig:interacting_data}(c)(i)], while the Thomas-Fermi limit is found at the top for $\Gamma \gtrapprox 1$ [Fig~\ref{fig:interacting_data}(c)(iii)]. 
Our numerical data points always trace the upper of the allowed channels throughout the entire range of self interaction strengths. 
\section{Physical Parameter Degeneracies and Observational Implications \label{Sec:degeneracy}}

\subsection{Physical Solitons and $m-\Gamma$ Degeneracy
\label{sec:degeneracy_1}}

\begin{figure}
    \centering
    \includegraphics[width=1\linewidth,keepaspectratio]{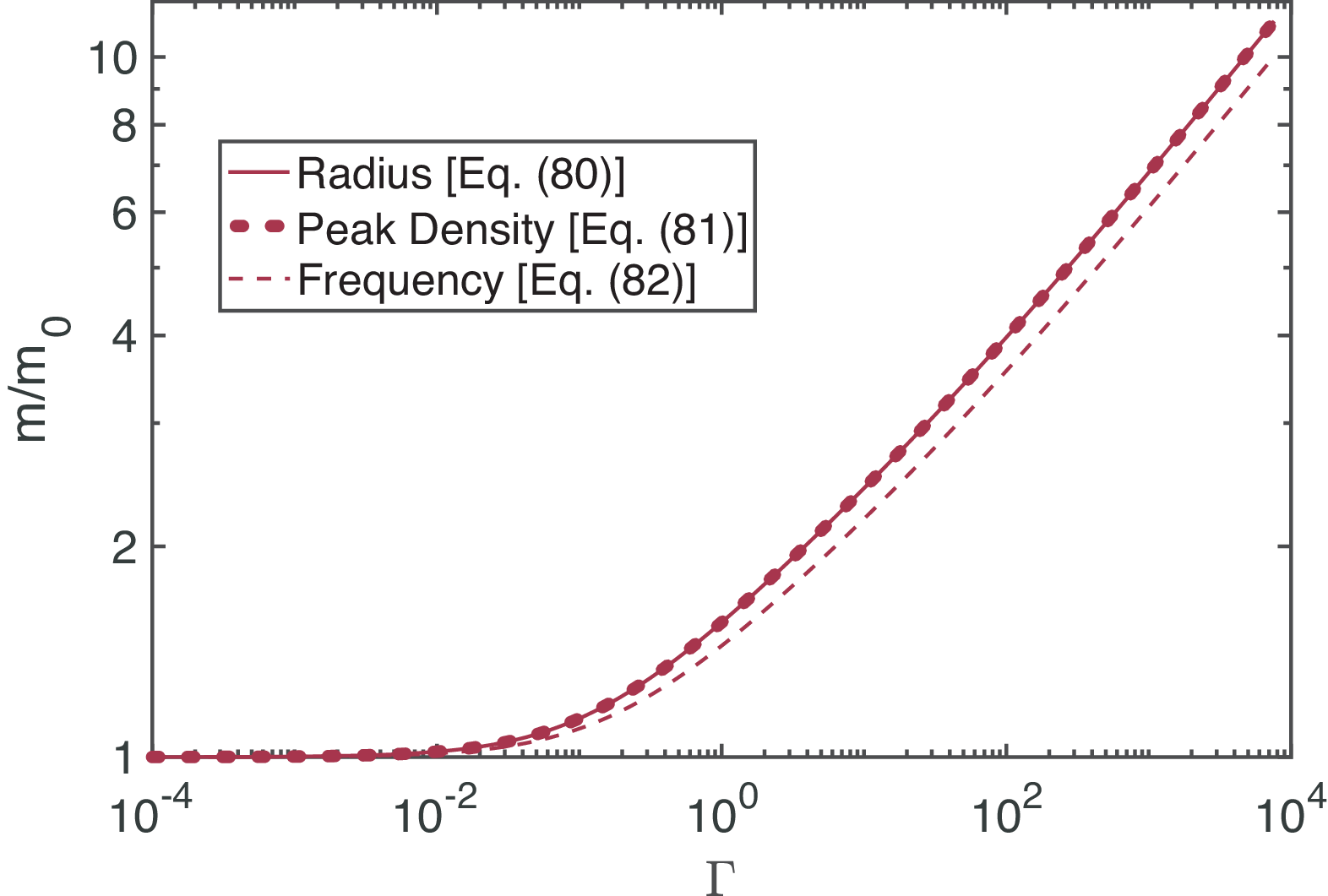}
    \caption{The scaled universal $m-\Gamma$ curves for which one obtains solitons of constant radius, peak density and oscillation frequency, corresponding to Eq.~(\ref{eq:r_degen}), Eq.~(\ref{eq:rho_degen}) and Eq.~(\ref{eq:f_degen}) respectively. Here, $m_0$ is defined as the boson mass at $\Gamma = 0$ for the respective soliton parameters $(r_c, f, \rho_0)$.}
    \label{fig:degen}
\end{figure}

Our preceding analysis for a soliton of fixed total mass $M$ has been conducted under an implicit further assumption of a fixed boson mass $m$. As anticipated, this has revealed a dependence of static and dynamical properties on the boson interaction parameter $g$, which enters as a new parameter appearing both implicitly and explicitly in the equations for the soliton radius $r_c(g)$, [Eq.~(\ref{eq:radius_g1})], the peak soliton density $\rho_0(g)$ [Eq.~(\ref{eq:rho_0_g1})] and the soliton oscillation frequency $f(g)$ [Eq.~(\ref{eq:frequency_g})]. 

To explain this, let us focus below on  the soliton radius $r_c$ in the absence or presence of interactions, for which we remind the reader of the main functional dependence of the previous expressions (under the assumption of a fixed $M$). These take the respective forms:
\\
For $g=0$:
\begin{equation}
    r_c(m) =  {\cal A}_0 \,\, \left( \frac{1}{m^2} \right) \;, 
\end{equation}
where
${\cal A}_0 = 2(\sigma_0/\nu_0)(\hbar^2/G M)$ depends on the non-interacting shape parameters (and the physical constants $\hbar$, $G$ and constant soliton mass $M$).
\\
For $g>0$:
\begin{align}
    r_c(m,\Gamma) &= {\cal A}(\Gamma) \left(\frac{1+\sqrt{1+ 15\mathcal{C}(\Gamma)} }{2}\right)\,\left(\frac{1}{m^2} \right) \label{eq:radius_mg} \\ \label{eq:rc_degen1}
    &= \mathcal{A}(\Gamma)\,\bar{\mathcal{C}}(\Gamma)\, \left(\frac{1}{m^2}\right)\;,
\end{align}
where
${\cal A}(\Gamma) = 2(\sigma(\Gamma)/\nu(\Gamma))(\hbar^2/G M)$ is defined in terms of the $\Gamma$-dependent shape parameters, and for convenience we have also introduced an expression $\bar{\mathcal{C}}(\Gamma) \ge 1$, which has convenient limiting cases 
\begin{align*}
\bar{\mathcal{C}}(\Gamma=0) &= 1, \\
\bar{\mathcal{C}}(\Gamma\rightarrow \infty)&= \frac{\sqrt{15\mathcal{C}(\Gamma)}}{2}\;.
\end{align*}

From Eq.~(\ref{eq:radius_mg}), we see that the inter-dependence of parameters $r_c$, $m$ and $\Gamma = g/g_*$ implies that it takes a combination of the values of two of these parameters to fix the third one.
Given that the main observationally-relevant physical quantities are likely to be the soliton mass $M$ (already assumed as constant in above discussion) and the soliton radius, we shall henceforth assume here a fixed value for the soliton radius $r_c(m,\Gamma)=R_c$, and consider the resulting inter-relation between the boson mass $m$ and the scaled boson self-interaction $\Gamma$. Thus by inverting Eq.~(\ref{eq:rc_degen1}), we obtain for each value of $R_c$ the following dependence of $m$ on $\Gamma$
\begin{equation}
     m(\Gamma) = \frac{1}{R_c^{1/2}}\left[{\cal A}(\Gamma)\,\bar{\mathcal{C}}(\Gamma)\right]^{1/2}  \label{eq:r_degen}.
\end{equation}
This expression shows clearly that fixing $r_c$ to the value $R_c$ and varying $\Gamma$ results in a varying $m(\Gamma)$ curve which becomes a locus of all values of the boson mass and dimensionless self-coupling corresponding to a soliton of fixed radius $R_c$. The idea of such curves corresponding to parameter values that result in the same soliton radius was first put forward in~\cite{ Chavanis2019,Chavanis:2020rdo}.

The same logic can be independently applied for fixed values of each of the peak density and the frequency. In particular, the corresponding equations defining these degeneracy curves read: \\
For a fixed peak density, $\rho_0$,
\begin{equation}\label{eq:rho_degen}
    m(\Gamma) = \left(\frac{ 4  \pi}{\rho_0} \right)^{1/6} \left[\left(\eta(\Gamma) \, \mathcal{A}(\Gamma)
 \right)^{1/6} (\bar{\mathcal{C}}(\Gamma))^{1/2}\right] \;.
\end{equation}
For fixed frequency, $f$,
\begin{eqnarray}
    m(\Gamma) &=& f^{1/3} \, \mathcal{B}(\Gamma) \left(\sqrt{\frac{2\bar{\mathcal{C}}(\Gamma)}{15\mathcal{C}(\Gamma)+2\bar{\mathcal{C}}(\Gamma)}} 
    \left(\bar{\mathcal{C}}(\Gamma)\right)^2\right)^{1/3} \label{eq:f_degen}
\end{eqnarray}
where we have introduced 
\begin{equation}
    \mathcal{B}(\Gamma) = \frac{\sqrt{\alpha(\Gamma)\sigma^3(\Gamma)}}{\nu^2(\Gamma)} \, \frac{4\sqrt{2}\pi\hbar^3 }{G^2M^2} .
\end{equation}
Equations (\ref{eq:rho_degen}) and (\ref{eq:f_degen}) determine the loci of all values of $m$ and $\Gamma$ which result in solitons of the same central density and same oscillation frequency respectively. In all cases, such \emph{degeneracy curves} reduce to their corresponding non-interacting limits when we take $\Gamma=0$, being consistent with a single viable boson mass value. 

Such dependences of the boson mass $m$ on $\Gamma$, resulting in $(m,\Gamma)$ pairs that support solitons of fixed radius, peak density and oscillation frequency, are shown in Fig.~\ref{fig:degen}.  Note that the boson mass has been scaled by a reference mass $m_0 = m(\Gamma=0)$ to make the plot universal. In this plot, we see clearly that the curves corresponding to fixed soliton radius\footnote{See similar figures 4 and 5 in \cite{Chavanis:2020rdo} for the case of solitons of fixed radius.} and fixed peak density overlap, whereas the scaling for fixed frequency reveals qualitative similarity but is distinctly identifiable.

\begin{figure}
    \centering
    \includegraphics[width=1\linewidth,keepaspectratio]{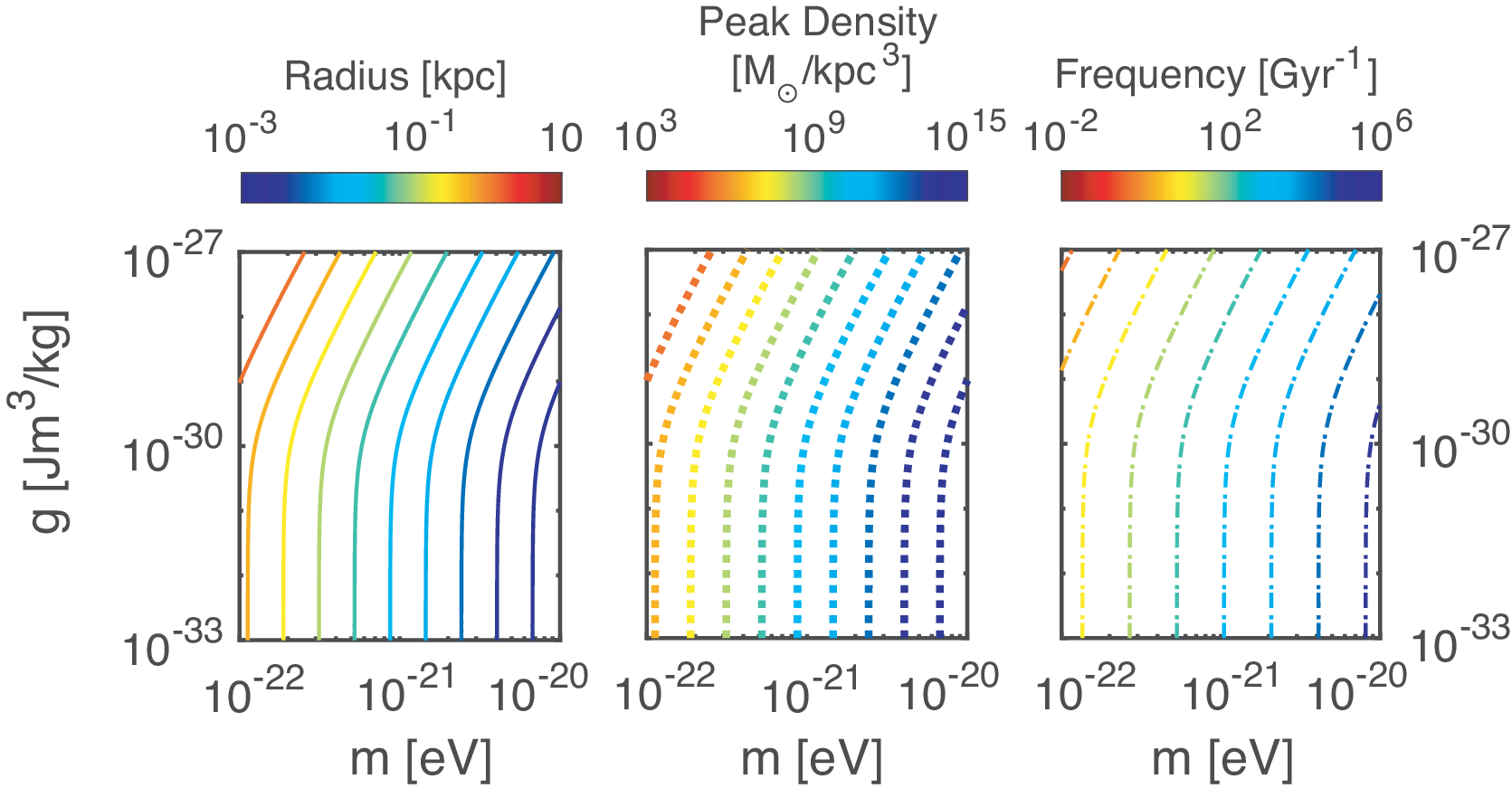}
    \caption{Curves of constant radius, peak density and oscillation frequency (the value of which is shown by the colourbars), revealing the allowed parameter space of repulsive self-interaction strength and boson mass based on such chosen parameters. These graphs can be extended towards the heavier boson mass regimes. The figure plotted for a soliton mass $M=10^8 M_\odot$.}
    \label{fig:contour}
\end{figure}

In the above discussion, the mass has been obtained as a function of the dimensionless interaction strength $\Gamma = g/g_*$ for fixed soliton radius, peak density or oscillation frequency. 
It can be instructive to reconsider the problem in terms of the relation between the boson self-interaction strength $g$, and the boson mass $m$, for any fixed value of soliton radius, peak density or oscillation frequency, parameterised through a particular value of $\Gamma$. This can be done by writing 
\begin{align}
    g(\Gamma,m) &= \Gamma \, g_*(m) \nonumber\\
    &= \Gamma \left(\frac{10\hbar^4}{G M^2}\right)\, \left(\frac{\sigma^2(\Gamma)}{\zeta(\Gamma)\nu(\Gamma)}\, \right) \, \frac{1}{m^3(\Gamma)}\;, \label{eq:g_gamma_m}
\end{align}
where $m(\Gamma)$ depends on $r_c$, $\rho_0$ or $f$, as given in  Eqs.~(\ref{eq:r_degen}), (\ref{eq:rho_degen}) and (\ref{eq:f_degen}) respectively. Tracing over the parametric variable $\Gamma$, and for each fixed value of radius, peak density or frequency, we can thus obtain a parametric relation between $g$ and $m$. Such dependence is plotted in the three panels of Fig.~\ref{fig:contour}, in which the fixed value of each such parameter is represented by a different colour line.

\begin{figure}
    \centering
    \includegraphics[width=1\linewidth,keepaspectratio]{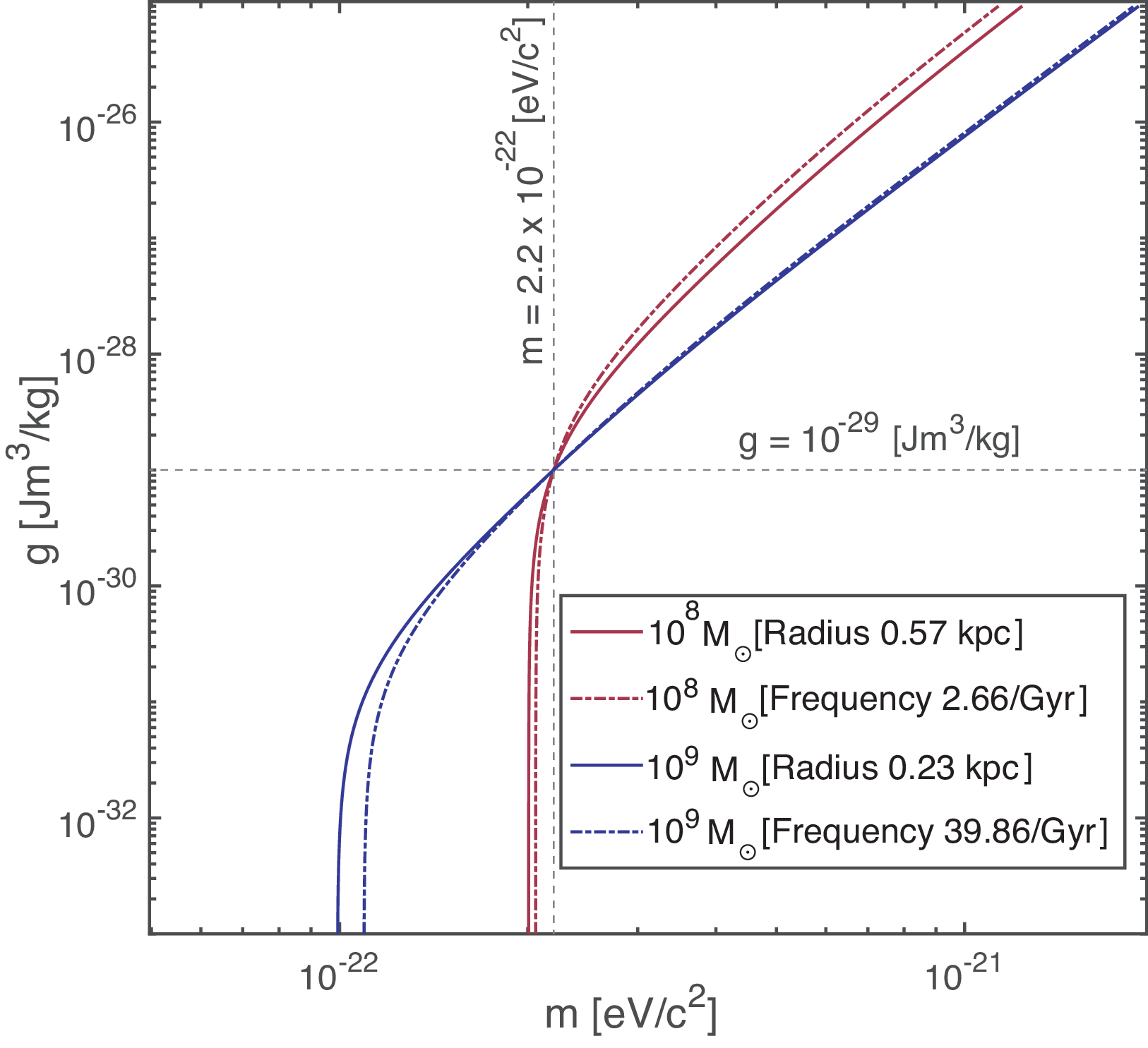}
    \caption{
    Scheme for simultaneous identification of boson self-interaction and mass through the numerical overlap of key soliton $g-m$ degeneracy lines for two different galaxies. Shown are heuristically chosen lines of constant radius (solid) and frequency (dashed) for soliton masses of $M=10^8 M_\odot$ (red) and $M=10^9 M_\odot$ (blue). Their overlap points to a unique combination of $g$ and $m$ values for such systems. 
    Here we have chosen to show a $0.57 \, \mathrm{kpc}$ radius and an oscillation frequency of $2.66 \, \mathrm{Gyr}^{-1}$ for the $10^8 M_\odot$ soliton, whereas the $10^9 M_\odot$ soliton has a $0.23 \, \mathrm{kpc}$ radius, and an oscillation frequency of $39.86 \, \mathrm{Gyr}^{-1}$. A single underlying FDM model would result in the degeneracy curves from a multitude of solitons with different masses, radii and oscillation frequencies all crossing at the same point (see text).   
    }
    \label{fig:intersect}
\end{figure}

\subsection{Observational Relevance
\label{sec:observations}}

For any single soliton with radius $r_c$ and peak density $\rho_0$, we have explicitly shown that there exists an entire contour line of boson mass and interaction strength pairings that support its existence within FDM. This rules out basing any constraints on a high-resolution accurate observation of a singular galactic dark matter object as the degeneracy means that no meaningful limits can be established; instead one can only deduce which contour the object must lie on. However, multiple objects of different total soliton mass will have differing contour lines, where the $g(m)$ gradients are dependent on their total mass. As a result, when one maps the contour lines of several galactic objects onto the same set of appropriately scaled axes, their intersection point will correspond to the true physical value of the boson mass and self-interaction strength which enables the formation of such solitons in our universe.

Fig.~\ref{fig:intersect} illustrates this point: First, we consider a specific pair of boson mass and self-interaction strength, $m = 2.2 \times 10^{-22} \, \mathrm{eV/c^2}$ and $g = 10^{-29} \, \mathrm{Jm^3/kg}$ which, for a given soliton mass, will yield a specific set of soliton parameters - the core radius, peak density and oscillation frequency. We then note that any points along the corresponding $m(g)$ degeneracy contour will yield an identical soliton - {\em i.e.} the resulting solitons are observationally indistinguishable for points chosen anywhere along a single degeneracy contour line for a fixed total mass. 

We can now design two solitons of different mass, chosen as $10^8 \, M_\odot$ and $10^9 \, M_\odot$, and plot their degeneracy contour lines in the $m\,-\,g$ parameter space, along which the radius and frequency remain constant (for fixed total mass). In this specific case, the $10^8 \, M_\odot$ soliton has a $0.57 \, \mathrm{kpc}$ radius and an oscillation frequency of $2.66 \, \mathrm{Gyr}^{-1}$, whereas the $10^9 \, M_\odot$ soliton has a $0.23 \, \mathrm{kpc}$ radius, and an oscillation frequency of $39.86 \, \mathrm{Gyr}^{-1}$. These values were chosen heuristically for the purpose of illustrating clearly how the information of the true boson mass and interaction strength may be extracted from the contour lines. The location of the intersect of these contour lines is the true boson mass and self-interaction strength which accommodates both of these solitons to exist in space. Once again, in our case the intersect values were arbitrarily chosen as an initialisation requirement. In a universe where the dark matter is fuzzy and described by a single scalar field, all degeneracy curves corresponding to candidate observed solitons must cross at the same point (within observational error bars) which would represent the single universal value for the mass $m$ and self-coupling $g$ that correspond to our universe. 

The existence of the aforementioned degeneracy could have an impact on current constraints on the boson mass which are likely to change if the parameter space is extended to include $g$.\footnote{Although the relative ranges of $m$ and $\Gamma$ exhibited in fig.~\ref{fig:degen} may imply that existing constraints on the boson mass will not be shifted by orders of magnitude, it is still the case that including the self-coupling may be relevant.} Indeed, various limitations have been placed upon the allowed range of the the boson mass by probing both cosmological scales \textit{e.g.}~\cite{Desjacques:2017fmf, Rogers:2020ltq}, as well as singular astronomical objects such as the star cluster orbiting the centre of Eridanus II \cite{Marsh2019, Chiang2021-2} or rotational curves and stellar kinematics~\cite{Bar2018, Maleki2020, Hernandez2023,Khelashvili2023}. Furthermore,~\cite{Alvarez-Rios2023} finds a correlation between the core oscillation and the oscillation of gas, which would further be affected by self-interaction and would prove a likely observational signature. All these works, apart from \cite{Desjacques:2017fmf}, do not include a self-interaction and in some cases clear differences between theoretical fits and observational data were identified~\cite{Hernandez2023}, while in others it was further evident that a single boson mass could not adequately fit theoretical curves to observational data~\cite{Bar2018, Khelashvili2023}. On the other hand,~\cite{Delgado2022} reports that a non-zero value of the self-coupling, along with a single boson mass, can fit the rotation curves of the dark matter dominated galaxies in the SPARC database, providing for the first time positive support for FDM solitons with a non-zero value of the self-interaction from rotation curves. The degeneracies discussed here would be relevant for all the above studies which make inferences about the boson mass.\footnote{In this work we discussed parameter degeneracies related to the compact FDM solitons. Similar degeneracies can also exist on cosmological scales for FDM overdensity power spectra, see \cite{Proukakis:2023txk} which analyses a linearized hydrodynamical version of the hybrid condensate particle model of~\cite{Proukakis:2023nmm}. } Clearly, all of the above are very preliminary observational investigations but provide strong motivation for precise, quantitative analysis of the model \cite{Indjin2024}. Steps have been taken to incorporate baryons into these models~\cite{Veltmaat2020-Baryons, Guzman2023}, which would likely add an additional layer of complexity. 

\section{Conclusions
\label{Sec:conclusions}}

In this paper we evaluated quantitatively the impact of a non-zero, repulsive self-interaction on the shape of virialized, fuzzy dark matter solitons and some of their key characteristics, namely their central density $\rho_0$, radius $r_c$ and the frequency $f$ of their small radial oscillations. We achieved this using a general ansatz for the density profile and quantified the change in its shape via the computation of 5 {\em shape parameters}, corresponding to dimensionless integrals that are associated to the different components of the soliton's energy, mass and moment of inertia. 

By using a dimensionless measure of the self-interaction, $\Gamma$, we mapped the transition of the shape parameters from the non-interacting to the strongly interacting regime with universal curves; all of the numerically generated solitons we explored (generated via imaginary time propagation with different masses $M$ and composed of bosons of different mass $m$), can be placed on these curves. Knowledge of the shape parameters then informs the transition of $\rho_0$, $r_c$ and $f$ from the non-interacting to the strongly interacting regimes. Again this transition can be described by universal functions if appropriately scaled quantities are used. As an interesting side-observation, we found that the accurate numerical determination of the shape parameters also allows for a better semi-analytical prediction of the oscillation frequency for $g=0$ compared to the use of either the Gaussian or empirical density profiles.   

Our results also led us to the notion of degeneracy curves, representing the loci of all those $(m,g)$ pairs that result in solitons with identical $r_c$, $\rho_0$ and $f$. The question then naturally arises of how such degeneracy might be broken via the observation of many possible solitonic objects. It would seem that such notions are highly topical given the increasing effort to place constraints on the parameters of FDM via confrontation with observations of astrophysical objects that could harbour FDM solitons. We leave further exploration of such confrontation to future work.

\section*{Acknowledgements}

I.K.L. acknowledges funding from European Union’s Horizon 2020 research and innovation programme under the Marie Sklodowska-Curie grant agreement No. 897324 (upgradeFDM), while G.R. and N.P.P. acknowledge funding from the Leverhulme Trust (Grant no. RPG-2021-010). We wish to thank P.-H.~Chavanis for comments on an earlier version of the manuscript and for bringing to our attention relevant aspects of his earlier works as well as the anonymous referee for pertinent comments that have allowed us to improve the clarity of the paper.  

\section*{Data Availability}
The data supporting this work are openly available at  \href{https://doi.org/10.25405/data.ncl.25592892.}{doi.org/10.25405/data.ncl.25592892.}.
\bibliography{Indjin2023_References.bib}

\bibliographystyle{iopart-num}

\end{document}